\documentclass[journal]{vgtc}              

\onlineid{1852}

\vgtccategory{Research}

\vgtcpapertype{theoretical \& empirical }

\title{A Preliminary Roadmap for LLMs as Assistants in\\Exploring, Analyzing, and Visualizing Knowledge Graphs}

\author{%
  \authororcid{Harry Li}{0000-0002-2288-6039},
  \authororcid{Gabriel Appleby}{0000-0003-2436-2121},  
  \authororcid{Ashley Suh}{0000-0001-6513-8447}
}

\authorfooter{
    \item Harry Li and Ashley Suh are with MIT Lincoln Laboratory. E-mail: \{harry.li, ashley.suh\}@ll.mit.edu.

  \item Gabriel Appleby is with Tufts University. E-mail: gabriel.appleby@tufts.edu.
}

\abstract{%
We present a mixed-methods study to explore how large language models (LLMs) can assist users in the visual exploration and analysis of knowledge graphs (KGs). We surveyed and interviewed 20 professionals from industry, government laboratories, and academia who regularly work with KGs and LLMs, either collaboratively or concurrently. Our findings show that participants overwhelmingly want an LLM to facilitate data retrieval from KGs through joint query construction, to identify interesting relationships in the KG through multi-turn conversation, and to create on-demand visualizations from the KG that enhance their trust in the LLM’s outputs. To interact with an LLM, participants strongly prefer a chat-based `widget,' built on top of their regular analysis workflows, with the ability to guide the LLM using their interactions with a visualization. When viewing an LLM's outputs, participants similarly prefer a combination of annotated visuals (e.g., subgraphs or tables extracted from the KG) alongside summarizing text. However, participants also expressed concerns about an LLM's ability to maintain semantic intent when translating natural language questions into KG queries, the risk of an LLM `hallucinating' false data from the KG, and the difficulties of engineering a ‘perfect prompt.’ From the analysis of our interviews, we contribute a preliminary roadmap for the design of LLM-driven knowledge graph exploration systems and outline future opportunities in this emergent design space.
}

\keywords{Large language models, knowledge graphs, visualization techniques and methodologies, human factors, human-AI collaboration, generative AI, visual communication}

\teaser{
  \centering
  \includegraphics[width=0.85\linewidth]{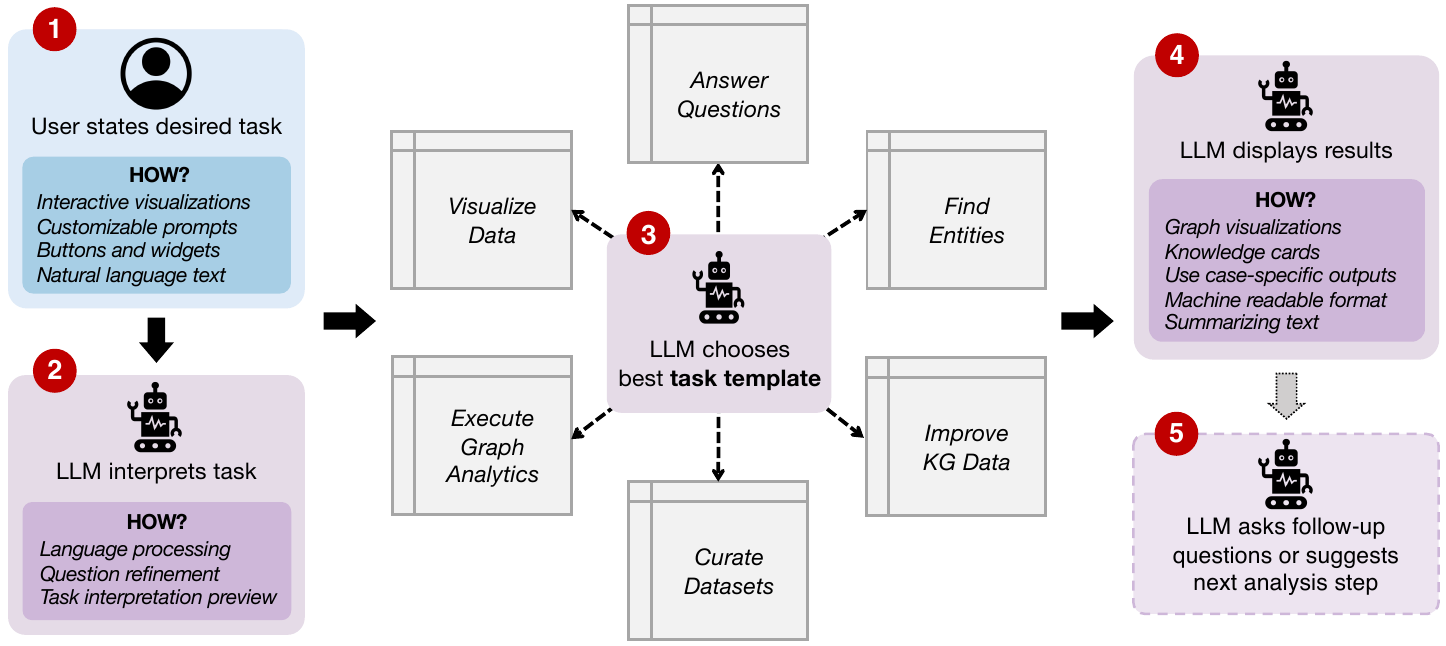}
    \caption{
    Exemplar workflow for an LLM assistant that supports users in exploring, analyzing, and visualizing the data in a KG, as derived from the findings of our mixed methods study (Section~\ref{sec:study}). (1) Users state their desired task through natural language, speech-to-text, customizable prompts, or interactions with a visualization. (2) The LLM interprets this task using natural language processing (NLP) techniques, initiating an iterative dialog to refine the user's task when necessary. Once the LLM is confident it has identified the user's goal, it (3) selects the corresponding \textit{task template} from the system to run. A template could contain code, precise instructions, or additional prompts for the LLM. (4) The LLM displays its results with a combination of visualizations and contextual text, then can optionally (5) continue this workflow by asking follow-up questions or suggesting a new task based on the analysis scenario. 
    We discuss how an LLM can act as an \textit{interpreter} and \textit{coordinator} for KG visual data analysis in Sections~\ref{sec:tasks},~\ref{sec:design-considerations}, and~\ref{sec:roadmap}.}
    \label{fig:workflow}
}

\graphicspath{{figures/}{figures/}{pictures/}{images/}{./}} 

\usepackage{lipsum}                    
\usepackage{tabularx}
\usepackage{tabulary}
\usepackage{capt-of}

\usepackage{csquotes}

\usepackage{color, colortbl}

\usepackage{makecell}

\usepackage{booktabs,siunitx}

\usepackage{lipsum}

\usepackage{enumitem}

\usepackage{mathptmx}                  

\usepackage{subcaption}

\usepackage{booktabs}
\usepackage{multirow}
\usepackage{graphicx}
\usepackage{soul}

\usepackage[normalem]{ulem}
\definecolor{mred}{rgb}{.80,.12,.30}
\definecolor{MRED}{rgb}{.80,.12,.30}
\definecolor{grey}{rgb}{0.5,0.5,0.5}
\definecolor{purple}{rgb}{.75,0,.85}
\definecolor{pistachio}{rgb}{0.58, 0.77, 0.45}
\definecolor{palesilver}{rgb}{0.9, 0.9, 0.9}

\newif\ifnotes
\notestrue

\let\origcite\cite
\renewcommand{\cite}[1]{\ifnotes\mbox{\origcite{#1}}\else \origcite{#1}\fi}

\begin{document}



\maketitle

\section{Introduction}


Recent advancements in large language models (LLMs) present an opportunity to assist users in visually exploring and analyzing knowledge graphs (KGs). KGs play a vital role across a variety of domains\cite{abu2021domain} by managing diverse data sources, organizing information semantically, and supporting complex machine learning tasks\cite{hong2020human}. Despite their ubiquity, KGs pose a number of challenges for users\cite{lissandrini2020graph, lissandrini2022knowledge, li2024kgs}, including difficulty in extracting meaningful insights, maintaining the accuracy of KG data, and creating compelling yet interpretable KG visualizations. In this paper, we work towards a preliminary roadmap for leveraging LLMs to lower the barrier of KG usage.

To inform this roadmap, we conducted a mixed methods study in which we surveyed and interviewed 20 KG and LLM experts from various fields, including industry, government laboratories, and academia. We asked participants a series of questions to identify how an LLM could best support their own tasks with KGs, their preferences for the interaction modality and presentation format of an LLM, and potential roadblocks they would expect to encounter when using an LLM assistant for their KG workflows. 
We performed a qualitative analysis\cite{macqueen1998codebook} to distill the most important themes from our participants' feedback, and use these findings to illuminate the goals, tasks, and visualization capabilities that users expect an LLM to facilitate.

Participants, particularly those who engage in KG analysis, expressed a strong interest in utilizing LLMs to refine their open-ended and domain-specific questions through multi-turn conversation until precise KG queries could be written. These participants also hoped an LLM would provide suggestions for what analysis steps to perform next, or recommendations for unexplored data of potential interest.  Participants who work on building KGs wanted help in improving the quality of their knowledge graphs (e.g., by the LLM deduplicating\cite{gal2014uncertain} semantically similar entities) and assistance with establishing new connections in the KG from disparate, text-based data sources. 

To communicate with the LLM, participants suggested a traditional chat-based or speech-to-text natural language interface (NLI) with prebuilt, customizable prompts. 
Interestingly, even LLM experts highly desired an interface with prebuilt prompts that ensure strong task performance. 
Participants also emphasized that they wanted interactive visualizations that could inform the LLM of their analysis intent. For example, when clicking on a visual element, prebuilt prompts or queries could be partially filled in 
amid back-and-forth dialogue. 
When receiving the outputs from an LLM, participants preferred a unified graphical representation of the results, verified from the KG, with a summary of those results written in natural language text. 

Participants cautioned us that integrating an LLM into their workflows may introduce incorrect information\cite{rawte2023survey}, as LLMs can impart biases and produce ``\textit{intentional lies}'' when returning results from KG data retrieval.
Moreover, using an LLM to answer questions from the KG may result in information loss either by misunderstanding the user's semantic intent, or by oversimplifying domain-specific concepts -- common limitations of NLIs\cite{setlur2016eviza}. Regardless, participants believed an LLM would likely perform better at these tasks than traditional approaches\cite{shen2022towards, karanikolas2023large}, so long as the user's expectations are aligned with the actual capabilities of an LLM\cite{zhang2023s}. 
As part of our roadmap, we offer strategies to address the concerns echoed by our study participants.

Finally, we propose design considerations for LLM-driven KG visual analysis systems and recommend potential workflows that can be implemented and improved upon, as illustrated in Figure~\ref{fig:workflow}. Specifically, we discuss step-by-step instructions on how systems can leverage LLMs as \textit{interpreters} and \textit{coordinators}, such that the LLM helps interpret a user's overarching goal, selects a \textit{task template} that best matches the derived task, and executes potentially many sequential analysis steps before returning a final result. This general framework lends itself to a number of specific applications within this emergent design space -- we provide an outline for promising research directions.  

To summarize, the major contributions of this paper are:

\begin{itemize}[topsep=1pt, partopsep=0pt,itemsep=2pt,parsep=0pt]
    \item A thematic analysis of surveys and interviews conducted with 20 KG and LLM experts across various fields and domains.
    \item A preliminary roadmap outlining the tasks that KG users want an LLM to facilitate, preferred input and response modalities for communicating with an LLM, and potential pitfalls that developers can anticipate in the design of future LLM and KG systems.
    \item Design considerations for LLM-driven KG visual analysis systems and detailed workflows to leverage conversational LLMs as assistants in exploring and visualizing KGs.
\end{itemize}

\section{Background \& Related Work}
Our related work covers a brief background on the historical challenges of using knowledge graphs, LLMs, and how they have been integrated in research to date.

\subsection{Knowledge Graphs and their Challenges}
The conceptual use of knowledge graphs dates back to the rise of the semantic web in 2001\cite{berners2001semantic} and gained prominence in 2012 after the creation of Google's Knowledge Graph\cite{ehrlinger2016towards}. KGs represent data as nodes (\textit{entities}), links or edges (\textit{relations}), and properties (\textit{attributes}). The semantic nature of KGs has made them ubiquitous for recommendation systems\cite{kepuska2018next}, explainable AI\cite{tiddi2020knowledge}, and language tasks -- particularly when combined with LLMs\cite{alkhamissi2022review, petroni2019language, brown2020language}.
For extensive details on KGs and their applications, see Hogan et al.'s survey\cite{hogan2021knowledge}.

Li et al.\cite{li2024kgs} identify the biggest obstacles to using knowledge graphs through an interview study with KG practitioners. First, poor data quality (e.g., missing, obsolete, or duplicate entries) makes maintaining the database problematic. KG \textit{Builders} struggle with performing manual but necessary updates to the KG, which is particularly burdensome if the KG scales to billions of entities, relations, and properties. 
The difficulties of managing a KG often lead to its deprecation, resulting in a loss of time and money for organizations. 

The biggest challenges faced by KG \textit{Analysts} include completing machine learning (ML) tasks and discovering relevant or interesting data. This is partly because writing queries is a time-consuming task (graph databases tend to have unique querying languages\cite{hong2020human}), and interface-based query builders do not provide feedback to users on why a query went wrong or how it can be modified to retrieve different data. Previous research\cite{ngonga2013sorry, Gan2021NaturalSM} aims to tackle this exact issue -- focusing on converting queries from and to natural language so users can understand the semantic meaning of their queries. We note, however, that many of these similarly proposed methods do not overcome the multitude of challenges for writing KG queries from scratch\cite{Bonatti:2019:Knowledge}.

Overly complicated and non-scalable KG visualizations\cite{klein2022bringing} were also cited as a major pain point in the workflows for KG users. Specifically, typical graph visualizations provided by KG systems\cite{lissandrini2020graph} were too complex for end-users, and tended to focus on irrelevant parts of the KG. Analysts also wanted the ability to perform visual sanity checking during their analysis of KG data, but the visualizations were either too information-dense, or difficult to generate due to the user's lack of visualization expertise.

In our paper, we begin to address how an LLM can be used to overcome these various problems. Specifically, we work with KG and LLM experts to understand how their KG practices could benefit (or be hindered) from the integration of an LLM. We offer a concrete set of tasks, interaction modalities, and interface designs that can inform the development of future LLM and KG systems. 

\subsection{Natural Language Interfaces and LLMs}
Natural language interfaces (NLIs) have become increasingly popular in the visualization community\cite{Aurisano2016Articulate2, Narechania2020NL4DVAT, mitra2022facilitating, huang2023flownl} -- in part because they help users explore, visualize, and model data without querying or database (DB) expertise\cite{li2014constructing}. However, many NLIs contributed in research suffer from limitations that an LLM could potentially overcome.

For example, Eviza\cite{setlur2016eviza} allows users to have a `dialog' with their DB by interactively refining and updating their queries with natural language. Its grammar-based approach, however, made it difficult to convert vague qualifiers like ``coldest'' or ``biggest'' into appropriate queries, and it was unable to recommend follow-up queries that a user might have. Snowy\cite{srinivasan2021snowy}, a template-based approach, addressed this limitation by suggesting utterances for visual analysis based on data `interestingness' metrics and pragmatics. While an improvement, a weakness of Snowy was its inability to infer the semantics of data grounded in the user's domain. The NLI Analyza\cite{dhamdhere2017analyza} specifically utilized knowledge graphs to overcome the problem of inferring semantics, but was limited by the scope of its synthetic KG.

NLIs have historically adopted ``grammar-translation'' approaches, relying on grammar and syntax-based rules. 
Interestingly, in 2021 Srinivasan and Setlur suggested the use of a chatbot to complement future iterations of the Snowy system\cite{srinivasan2021snowy}. 
Karanikolas et al.\cite{karanikolas2023large} provide a thorough background on the differences in technologies using LLMs versus classical Natural Language Understanding (NLU) and Natural Language Generation (NLG) approaches.

Today, research in NLIs and intelligent user interfaces (IUIs) investigates how LLMs can overcome a multitude of unique visual analysis challenges. 
LLMs have been used to generate data insights from computational notebooks\cite{lin2024inksight}, create animated data videos from visualizations\cite{shen2024dataplayer}, communicate visual tasks to robots\cite{zeng2023large}, help developers visually understand their code\cite{nam2024using}, and create custom charts from visualization libraries\cite{vazquez2024llms}. Similar in motivation to our work, Sultanum and Srinivasan contributed DataTales\cite{sultanum2023datatales}, a visual story authoring tool showcasing how LLMs can aid in data understanding during exploratory analysis. After a chart is created, an LLM provides an on-demand narrative telling a story about the visualized data. 

Our paper seeks to broadly understand how an LLM could similarly aid users in the visual data exploration and analysis of KGs. There are a multitude of benefits to using LLMs in this space -- we discuss them in detail in Section~\ref{sec:tasks}. LLMs can enable a more nuanced interpretation of a user's question for a knowledge graph, a sensible feature given the semantic richness of KGs\cite{Bonatti:2019:Knowledge}. Similarly, LLMs can help refine a user's complex, vague, or multi-part question into precise KG queries\cite{yang2023llm} with the appropriate conditions, filters, or criteria. We posit that, while some LLMs may be limited at performing some KG tasks, a multi-agent pipeline\cite{wu2023autogen} could be implemented to help overcome this. We discuss further in Section~\ref{sec:roadmap}.

\subsection{Integration of LLMs with Knowledge Graphs}
For a review on the integration of LLMs with structured knowledge sources (e.g., KGs and ontologies), see\cite{petroni2019language, alkhamissi2022review}.

There is a large body of work investigating how to simultaneously use LLMs and KGs for data-driven tasks. Pan et al. recently published their own roadmap that categories three major research areas: 1) \textit{KG-enhanced LLMs}, which use KGs to improve LLM training and inference 2) \textit{LLM-augmented KGs}, which use LLMs to support KG-based tasks, and 3) \textit{Synergized LLMs \& KGs}, where they play equal roles in knowledge representation and reasoning\cite{2024_unifying_llms_and_kgs}. Our study focuses primarily on the second category -- how LLMs can assist in KG analytic tasks. While Pan et al. distill how LLMs can help \textit{architecture} a KG, we instead focus on investigating how LLMs can be used to support practitioners in their KG exploration, analysis, and visualization tasks.

Recent work has suggested that the integration of KGs can address the shortcomings of LLMs. A popular line of research in this space is the use of KGs to mitigate `\textit{hallucinations}' from LLMs\cite{martino2023knowledge, agrawal2023can}, that is, an LLM's ability to generate factually untrue or fabricated information\cite{rawte2023survey}.
To address these weaknesses, most research
focuses on techniques where the LLM reasons (i.e. traverses) over a KG to retrieve up-to-date facts in the data\cite{sun2023think, wen2023mindmap, luo2023reasoning, knowledgeInjecitonToCounter, sen-etal-2023-knowledge} -- essentially forcing the LLM to fact-check itself against the ground truth of a KG. 

While these works highlight the capabilities of integrating KGs in the design and development of LLMs, our study focuses on the other direction of this integration: how an LLM can be integrated to lower the barrier of KG usage. In Section~\ref{sec:discussion}, we discuss research opportunities for the other direction: using a KG to improve LLM-driven NLIs.  
\section{Methodology}
\label{sec:study}

\begin{table*}[h!]
\centering
\renewcommand{\arraystretch}{1.4}
\centering
\resizebox{.95\linewidth}{!}{%
\definecolor{palesilver}{rgb}{0.9, 0.9, 0.9}

\definecolor{none}{RGB}{217,217,217}
\definecolor{slight}{RGB}{189,189,189}
\definecolor{some}{RGB}{150,150,150}
\definecolor{moderate}{RGB}{99,99,99}
\definecolor{extreme}{RGB}{37,37,37}


\renewcommand\theadalign{bt}
\renewcommand\theadfont{\bfseries}
\renewcommand\theadgape{\Gape[4pt]}
\renewcommand\cellgape{\Gape[4pt]}

\sffamily
\resizebox{\textwidth}{!}{
\begin{tabular}[t]{lllllllll}
\toprule

\textbf{PID} & 
\textbf{Education} & 
\textbf{Affiliation} & 
\textbf{Job Title} & 
\textbf{Primary KG Use Case} & 
\textbf{KG Persona(s)} &
\thead{Familiarity\\with KGs} &
\thead{Familiarity\\with LLMs} &
\thead{KG+LLM\\Usage}\\
\midrule

1 & MS & FFRDC & Research Scientist & Network Analysis & Builder, Analyst & 3 \textcolor{some}{(Some)} & 4 \textcolor{moderate}{(Moderate)} & Together\\ 
2 & MS & FFRDC & Research Scientist & Text Analysis & Builder, Analyst & 3 \textcolor{some}{(Some)} & 3 \textcolor{some}{(Some)} & Separately\\ 
3 & BS & FFRDC & Research Scientist & Text Analysis & Analyst & 3 \textcolor{some}{(Some)} & 5 \textcolor{extreme}{(Extreme)} & Together\\ 
4 & MS & FFRDC & Research Scientist & Network Analysis & Builder, Analyst & 4 \textcolor{moderate}{(Moderate)} & 3 \textcolor{some}{(Some)} & Together\\ 
5 & PhD & FFRDC & Research Scientist & Text \& Network Analysis & Builder, Analyst & 3 \textcolor{some}{(Some)} & 4 \textcolor{moderate}{(Moderate)} & Together\\ 
6 & PhD & FFRDC & Research Scientist & Network Analysis & Builder, Analyst & 3 \textcolor{some}{(Some)} & 2 \textcolor{slight}{(Slight)} & Together\\ 
7 & MS & FFRDC & Research Scientist & Social Network Analysis & Builder, Analyst & 4 \textcolor{moderate}{(Moderate)} & 3 \textcolor{some}{(Some)} & Separately\\ 
8 & MS & FFRDC & Research Scientist & Anomaly Detection & Builder, Analyst & 2 \textcolor{slight}{(Slight)} & 5 \textcolor{extreme}{(Extreme)} & Separately\\ 
9 & MS & FFRDC & Research Scientist & User Workflow Analysis & Analyst & 3 \textcolor{some}{(Some)} & 4 \textcolor{moderate}{(Moderate)} & Separately\\ 
10 & MS & FFRDC & Research Scientist & Social Network Analysis & Builder, Analyst & 3 \textcolor{some}{(Some)} & 3 \textcolor{some}{(Some)} & Separately\\ 
11 & PhD & FFRDC & Research Scientist & Network Analysis & Analyst & 3 \textcolor{some}{(Some)} & 4 \textcolor{moderate}{(Moderate)} & Separately\\ 
12 & MS & FFRDC & Research Scientist & Network Analysis & Analyst & 2 \textcolor{slight}{(Slight)} & 3 \textcolor{some}{(Some)} & Separately\\ 
13 & PhD & Industry & Data Scientist & Drug Discovery & Builder, Analyst & 5 \textcolor{extreme}{(Extreme)} & 2 \textcolor{slight}{(Slight)} & Together\\ 
14 & PhD & Industry & Data Scientist & Drug Discovery & Builder, Analyst & 3 \textcolor{some}{(Some)} & 3 \textcolor{some}{(Some)} & Together\\ 
15 & MS & Industry & Data Scientist & Drug Discovery & Analyst & 2 \textcolor{slight}{(Slight)} & 4 \textcolor{moderate}{(Moderate)} & Together\\ 
16 & MBA & Industry & Tech. Product Director & Drug Discovery & Consumer & 3 \textcolor{some}{(Some)} & 3 \textcolor{some}{(Some)} & Separately\\ 
17 & PhD & Academia & Faculty & Text Analysis & Analyst & 2 \textcolor{slight}{(Slight)} & 5 \textcolor{extreme}{(Extreme)} & Separately\\ 
18 & MS & Industry, Academia & Data Scientist, Student & Stock Prediction & Builder, Analyst & 4 \textcolor{moderate}{(Moderate)} & 2 \textcolor{slight}{(Slight)} & Separately\\ 
19 & PhD & Academia & Faculty & Data Augmentation & Analyst & 4 \textcolor{moderate}{(Moderate)} & 2 \textcolor{slight}{(Slight)} & Separately\\ 
20 & PhD & Academia & Faculty & Text Analysis & Builder & 3 \textcolor{some}{(Some)} & 4 \textcolor{moderate}{(Moderate)} & Together \\

\bottomrule

\end{tabular}}}
\caption{Demographics for the participants of our study, described in Section~\ref{sec:study}. From left to right: the participant's ID; highest education; organization they work for (\textit{FFRDC} stands for Federally Funded Research and Development Center); organizational role; main use cases for KGs; KG persona characterization\cite{li2024kgs}; overall familiarity working with KGs and LLMs; and whether they typically use KGs and LLMs together or separately. Familiarity was self-reported by our participants on a Likert Scale from (1) \textit{not at all familiar} to (5) \textit{extremely familiar}.} 
\label{table:participants}
\end{table*}

We conducted a mixed methods study that included an open-response survey\cite{kintzer1977advantages}, pair-interviews\cite{akbaba2023two}, and focus group interviews\cite{rabiee2004focus}. 

\subsection{Study Goal}
The purpose of our study was to inform a preliminary roadmap for using LLMs as assistants in KG exploration and analysis. To do so, we focused on three overarching topics: the \textbf{tasks} that LLMs can best facilitate for users, the preferred \textbf{modalities} for communicating with or receiving results from an LLM, and the possible \textbf{pitfalls} that designers might face when integrating LLMs with KGs for visual data analysis.

Our mixed methods approach\cite{curry2009qualitative} was intended to reach as many participants as possible for broad feedback on this potential design space. For each method of response collection (i.e.\ open-response surveys and interviews), we asked participants the same set of questions. In some cases, we held follow-up calls with survey participants to better understand the responses we received from them. 

\subsection{Participants}
\label{sec:participants}
Participants were recruited from a combination of professional mailing lists and snowball sampling\cite{naderifar2017snowball}. 
We required that participants be familiar with both knowledge graphs and LLMs, and have expertise with at least one.
A complete list of our study participants, along with a brief description of their demographics, is shown in Table~\ref{table:participants}. 

Participants self-reported their familiarity with knowledge graphs and LLMs using a Likert Scale of 1 (\textit{No Familiarity}) through 5 (\textit{Extreme Familiarity}). We classified participants as either KG Builders (those who create KGs), Analysts (those who generate insights from KGs), or Consumers (those who use KG-based systems but not KGs directly) using Li et al.'s KG persona characterization\cite{li2024kgs}. 

Participants P1-P12 come from the same federally funded research and development center (FFRDC) but work across three different divisions focusing on varying problem domains. Participants P13-P16 come from two different enterprise organizations, P17 is an industry data scientist and PhD student, and P18-P20 are faculty members at different universities. 


\subsection{Protocol}
All interviews were conducted by two of the authors, following the pair-interview guidelines in\cite{akbaba2023two}. Immediately after each interview, both authors recapped the discussion together and compared notes. We further discuss our qualitative analysis in Section~\ref{sec:analysis}. 

Our open-response survey was created using Microsoft Forms. Participants P1-P10, P13-P14, and P16-P19 completed the open-response survey. Follow-up interviews were conducted with P1 (\textit{Pair-Interview}), P4 \& P5 (\textit{Focus Group \#1}), P14 (\textit{Pair-Interview}), and P13 \& P15 (\textit{Focus Group \#2}). Focus group \#2 was held in person and lasted roughly two hours. The remaining interviews were held over Zoom for an hour. In each of these follow-ups, we asked participants to (1) elaborate on their survey responses and (2) provide feedback on possible interface designs.
Participants were asked whether proposed designs matched their expectations, and if there were any possible improvements or missing features. We discuss this feedback in Sections~\ref{sec:llm-justification},~\ref{sec:output-modality}, and~\ref{sec:roadmap}.

Three participants did not take the open-response survey and instead answered the same set of questions with us verbally over Zoom: P11 \& P12 (\textit{Focus Group \#3)}, and P20 (\textit{Pair Interview}).  Each interview was semi-structured and lasted one hour. We were connected to P11, P12, and P20 through snowball sampling and consider them to be primarily LLM experts. 
Therefore, we focused on eliciting their detailed feedback on an LLM's potential capabilities as an analysis assistant. 
In total, we interviewed 9/20 participants either in-person or over Zoom.

\subsection{Questionnaire}

The questionnaire presented to participants was consistent across our distributed survey and conducted interviews. For the open-response survey, we presented each question with a free-form textbox. For our semi-structured interviews, we walked through a slide deck where questions were presented one slide at a time. 

At the start of the questionnaire, we included a high-level description of our research goal: ``\textit{We want to build a visual interface with an LLM facilitator to help users explore the data in their KGs.}'' We then explained that to determine design goals, we were eliciting feedback on what types of tasks, questions, and modalities participants believed this interface would best support, given their own workflows.

The four questions presented to participants were:

\begin{enumerate}[topsep=2pt, partopsep=0pt,itemsep=1pt,parsep=2pt]
  \item \textbf{Tasks:} Assume you are shown an interface that lets you explore a KG with the help of an LLM assistant. What tasks, goals, or questions would you expect the LLM to facilitate for you? 
  \item \textbf{Interaction Modality:} How would you prefer to interact or communicate with the LLM when exploring the KG? 
  \item \textbf{Presentation Modality:} How would you prefer the LLM and/or KG to present its outputs to you during exploration?
  \item \textbf{Potential Pitfalls:} Are there areas you believe the LLM would be particularly good or bad at for you when exploring the KG? 
\end{enumerate}

Participants were also able to provide any other high-level thoughts or opinions on the use of an LLM assistant for KG exploration. We include this feedback, where appropriate, in the entirety of our results. 

\subsection{Analysis}
\label{sec:analysis}

The goal of our analysis was to exhaustively identify all ways that an LLM could benefit or hinder our participants' KG workflows. 
We performed qualitative coding following a thematic analysis approach\cite{braun2006using} to distill the results of our survey and interview data. 
Both interviewers (authors) individually coded each participant's response to the questionnaire, where a code was treated as an overarching but concise description of an individual response\cite{decuir2011developing}. An individual code could only be applied once per response (or utterance), but many codes could be assigned when applicable.

While coding responses, we found that participants would occasionally reference their previous answers -- we did not double-count codes in these cases. If a response included additional information related to a previous question, we assigned new codes accordingly. Our interview data was coded in the same manner, where a participant's response to a question was treated as a single utterance in conversation. When interview participants would similarly bring up a new idea related to a previous question, we coded it as a separate utterance. 

After coding was complete, all three authors met to merge similar tags and collaboratively resolve any disagreements. While analyzing the response data, we also examined code counts across all participants for each question. We include a full table of our coded responses, per participant and question, as supplemental material.

In the presentation of our findings, we include code counts for each question to highlight the distribution of responses we received. The results of our study form the building blocks for our proposed roadmap.
\section{LLM Tasks for KG Exploration \& Analysis}
\label{sec:tasks}

\begin{figure}[h!]
    \centering
    \includegraphics[width=0.99\linewidth]{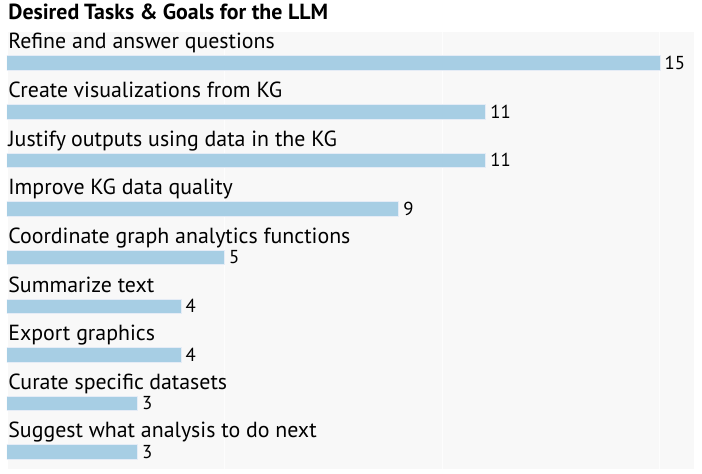}
    \caption{Distribution of responses from participants on which tasks or analytic goals they want an LLM to help facilitate when using a KG. We discuss each of these tasks in Section~\ref{sec:tasks}.}
    \label{fig:q1}
\end{figure}

We asked participants to describe the goals, tasks, and types of questions they would want an LLM assistant to facilitate. All coded responses are shown in Figure~\ref{fig:q1}. The tasks we distill below serve as a starting point for the design of future LLM-driven KG exploration systems.

\subsection{Answer Questions}
\label{sec:qa-task}
Assistance with question-answering from the KG was by far the most common task requested from participants (15/20). This task encapsulates help with creating and refining KG queries, finding entities given a set of constraints (e.g., ``find all nodes with \textit{this} kind of property''), summarizing the relationships between entities, and identifying paths or multi-connections in the KG. While this task is similar in nature to a data retrieval task, we note that most participants discussed the LLM gathering data for them through question-answering. 

\subsubsection{Convert Natural Language to KG Queries}
Rather than manually crafting queries on their own -- which can be difficult due to the number of unique KG querying languages\cite{li2024kgs} -- participants asked for the LLM to help them generate syntactically correct queries from their natural language questions:

   \begin{displayquote}
        \textit{I like that LLMs could be used in situations normally reserved for a technical specialist. If an LLM could help me write queries, I wouldn't have to memorize Cypher\cite{francis2018cypher}.} -P11
    \end{displayquote}

Even users who have proficiency with KG querying languages said an LLM could still help them overcome this burdensome, time-consuming task:

\begin{displayquote}
    \textit{I've always found KGs to be very cumbersome \ldots they're a time sink. If there's a more straightforward way to get to the answer I'm trying to find, that'd be great.} -P14
\end{displayquote}

\begin{displayquote}
    \textit{I think an LLM assistant could play a major role in semi-automating KG search capabilities. Some of the KG search strings are quite convoluted. Having an assistant formulate those commands for you to act as a go-between
    \ldots 
    would be very helpful as an analyst.} -P9
\end{displayquote}

However, participants still wanted the ability to customize the LLM's translated query before executing it on the KG: ``\textit{it would be very helpful if I could edit the query language instructions afterwards}'' (P17).

LLMs designed to write code, e.g., Code Llama\cite{rozière2024code}, could be potentially used as a starting point for these question-answering tasks. Work by Yang et al.\cite{yang2023llm} begins to tackle the development of SPARQL-writing LLMs for knowledge-based question answering\cite{zhang2021neural}.

\subsubsection{Refine Questions through Multi-Turn Conversation}
\label{sec:task-refine}
Four participants highlighted that an advantage of LLMs is their ability to engage in an iterative dialogue to refine the user's analysis questions:

\begin{displayquote}
    \textit{LLMs help people engage in multi-rounds and back-and-forths, which is not necessarily the case when sending a query to a KG. Multi-turn dialogue allows you to clarify your questions, revise your questions, etc. This is an advantage [of using LLMs].} -P20
\end{displayquote}



A back-and-forth dialogue between an LLM and user can help facilitate analysis refinement, where a user's open-ended exploratory (or domain-specific) questions could be further specified through conversation (e.g., ``\textit{what do you mean by} interesting \textit{and} unusual\textit{?}''). We note that this interaction is similar to the behavior of conversational agents.

\begin{displayquote}
    \textit{Rather than re-write and re-submit a query, an iterative refinement approach would be useful, where you can tell the LLM assistant `that was good, but I want more information about X.'} -P3
\end{displayquote}

After executing a query, four participants wanted the LLM to summarize the results in a way that could improve their own data comprehension: ``\textit{Once I had the right question, I would expect the LLM to retrieve all relevant data from the KG and explain it to me}'' (P19). 
%

\subsection{Create Visualizations from the KG}
\label{sec:justify}
Eleven participants wanted an LLM assistant to supply on-demand visualizations from the KG throughout any conversation. This was in part to help visually summarize large amounts of text or data (4/20), and to justify the outputs of the LLM (11/20). 

It is important to note that \textbf{create visualizations} was often discussed as a \textit{means} for the LLM to complete the tasks outlined in this section -- in other words, visualization is the \textit{How} in Brehmer and Munzner's task typology\cite{brehmer2013multi}. For the sake of clarity, we discuss the most common reason that participants wanted an LLM to create a visualization for them: ``\textit{to increase trust in the LLM.}'' 

\subsubsection{Justify the LLM's Outputs}
\label{sec:llm-justification}
Participants expressed they would like the LLM to justify its outputs using data from the KG, specifically to help guard against its hallucinations\cite{agrawal2023can}: ``\textit{ Ideally these tools will point back to the underlying knowledge graph, rather than using it behind the scenes}'' (P17).

Methods for justifying the outputs of an LLM -- other than supplemental graph visualizations -- include hyperlinks that point to relevant data in the KG, or any source documents that were used to generate the KG's data. 
P4 suggested ``\textit{a model that could tell users - this information in the output comes from the KG, the other information comes from the LLM's own knowledge.}'' 

Participants described an interface that was designed to connect the generated output from the LLM to partial visualizations of the KG:

\begin{displayquote}
    \textit{I would want to always see what parts of the LLM's response was drawn from the KG. I don't want to see an LLM's hallucinations and believe that it is factual and drawn from my KG. I would like to see some sort of linkages between words in the LLM's response and actual nodes and links in the KG.} -P19
\end{displayquote}   

P14 told us that his data science team ``\textit{always takes the LLM's outputs with a grain of salt},'' and will often double-check their own sources ``\textit{to see if what the LLM told us was true.}'' 
When we asked how this workflow could be improved, he told us that facts within the KG -- text, nodes, or links -- should be visually represented on screen when referenced by the LLM. Moreover, if the LLM cites a research paper or article, the paper's title and abstract should always be provided so they can skim the content themselves. 

\begin{figure}[ht!]
    \centering
    \includegraphics[clip, trim=0cm 0 6.5cm 0, width=1\linewidth]{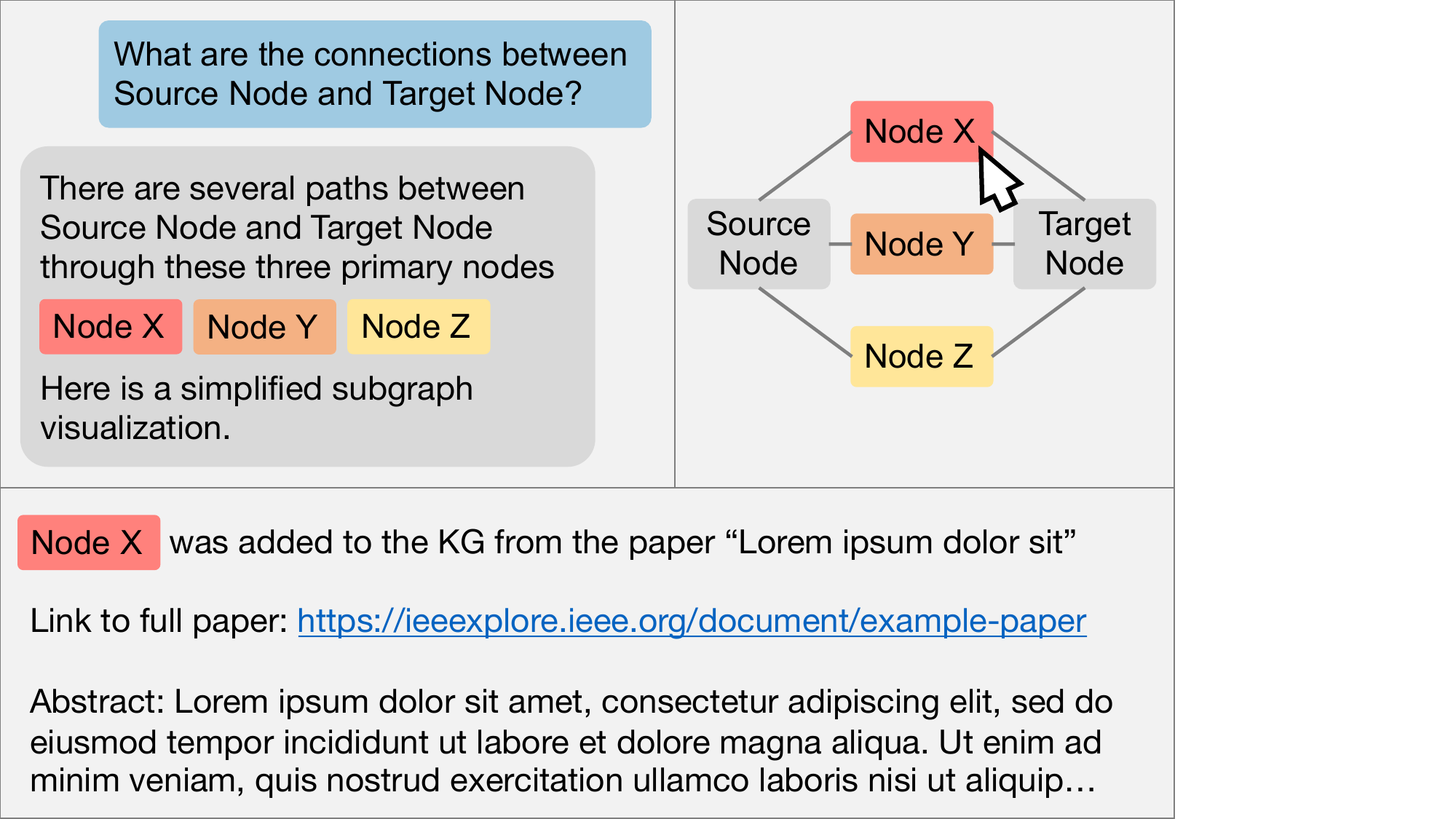} 
    \caption{A wireframe design, informed by participants' responses (Section~\ref{sec:justify}), showing an LLM's outputs with built-in justification. On the left side, the user asks about connections between two nodes in the KG. The LLM responds with a list of target entities that connect the two nodes. On the right, the subgraph visualization from the KG shows the simplified paths between the nodes. When the user clicks on \textit{Node X}, the panel on the bottom shows the original data source. Participants remarked that this type of design could help them examine true connections in the source data, thus making them more aware of the validity of an LLM's responses. We discuss the workflow for this system in Section~\ref{sec:workflow}.}
    \label{fig:llm-justification-wireframe}
\end{figure}

An illustrative example of this interface, based on our participants' feedback, is shown in Figure~\ref{fig:llm-justification-wireframe}.

\subsection{Improve and Simplify KG Data} 
\label{sec:task-improve}
Nine participants requested the ability for an LLM to help them validate the quality of the data in their KGs. This includes identifying conflicting or obsolete data\cite{lissandrini2022knowledge}, and finding duplicate entries in the KG with the same meaning but different labels (i.e. perform deduplication\cite{gal2014uncertain}): 

\begin{displayquote}
    \textit{A disease could be called different things, and the LLM can disambiguate them easily. The LLM does a good job at harmonizing different names [in the KG].}'' -P14
\end{displayquote}

LLMs could also be used to verify whether relationships in the KG make logical sense\cite{paulheim2017knowledge}: 

\begin{displayquote}
    \textit{Another interesting task would be KG verification - given a target node, do all of its links make sense? For example, if the target node was `apple' and the links were `is a fruit', `is edible,' and `is purple,' my expectation is that the LLM would be able to identify that the last link didn't belong and highlight it for my review.} -P8
\end{displayquote}

\subsubsection{Build and Complete KGs}
\label{sec:task-createKG}
When creating KGs, participants wanted the LLM to help predict nodes or links that are missing\cite{toussaint2022troubles} -- a common machine learning and graph neural network task\cite{chen2020knowledge}. 

Two of our participants explicitly mentioned that, because LLMs can reason over natural language text, they can also be used to generate new data (from text-based features) for the KG: ``\textit{An LLM would be good at processing text-based features in a KG if most data associated with the nodes or edges is stored in a text-based format}'' (P5).

Similarly, P7 noted that an LLM could help him ``\textit{augment the knowledge graph}'' by ``\textit{identifying topics that are semantically similar but might not be already associated with each other in the knowledge graph.}'' Unlike the simplification of redundant KG data, this task would allow the LLM to build new edges in the graph. 

Yao et al.\cite{yao2019kg} demonstrate how pre-trained language models can be successfully used for knowledge graph completion. 

\subsection{Execute Graph Analysis Functions}
Two of our participants mentioned that LLMs are not well suited to calculate traditional graph analysis metrics like centrality, diameter, or shortest path. Graph database query languages (e.g., SPARQL) have been historically criticized for their inability to execute these types of algorithms, particularly when aggregating a user's desired data\cite{Bonatti:2019:Knowledge}.

Regardless, five participants recommended that different graph functions (or toolkits) could be supplied to an LLM for it to coordinate and use, building from the user's analysis question:

\begin{displayquote}
    \textit{If the LLM can be used as an interface that translates natural language intent to an underlying graph analysis tool like NetworkX, it could calculate metrics like centrality.} -P12
\end{displayquote}

P1 suggested a similar workflow, ``\textit{if you have a collection of analysis tools or routines, maybe you could use the LLM to write programs to coordinate them.}'' We discuss the LLM acting as an analysis coordinator further in Section~\ref{sec:workflow}.

\subsection{Export User-Defined Data}
Seven participants wanted the LLM to help them export KG data.

\subsubsection{Design Graphics for Presentations and Reports}
\label{sec:task-make-graphics}
Four participants wanted the LLM to help them create and export graphics to insert into presentations or reports:

\begin{displayquote}
    \textit{After running a query, it would be nice if the LLM could generate an image for you\ldots when we make slides for presentations, the tedious part is always having to create graphics and pictures.} -P20
\end{displayquote}   

Generating data science presentations and reports is a historically tedious task\cite{zheng2022telling}.
P16 told us that ``\textit{ready-to-use slide decks are a need}'' when asked what the LLM should produce from a knowledge graph.

A pipeline in which an LLM is coordinating appropriate extraction from the KG for visualization creation (e.g., writing or executing D3 code\cite{chen2023beyond}) could support this task. Vazquez\cite{vazquez2024llms} evaluates how well LLMs can coordinate and create visualizations from common chart-building toolkits, like Plotly. Alternatively, LLMs designed to generate images from text (e.g.,\cite{koh2024generating}) could also be used.

\subsubsection{Curate Custom Datasets}
Three participants wanted the LLM to ``\textit{extract datasets from the KG given some constraints,}'' describing it as a ``\textit{really powerful feature if the LLM is robust enough}'' (P10). However, P10 warned that downstream processing could be greatly impacted if the LLM introduced errors in the produced dataset. 

Systems like CAVA\cite{Cashman:2020:CAVA}, which support users in augmenting their datasets by crawling knowledge graphs, could be extended to include an LLM for this task. Specifically, an LLM could be used to semantically `search' the knowledge graph for additional data, identify data with similar labels, or find data with disparate disconnections -- all challenges that are difficult to overcome with the KG alone\cite{lissandrini2022knowledge}. 

\subsection{Make Recommendations}

Three participants told us that the LLM should be able to make recommendations. This included what types of analyses to perform, and ``\textit{what related data should be explored next based on the initial prompts}'' (P9). Interestingly, all three participants who requested this feature were primarily LLM experts rather than KG experts -- suggesting that these recommendations may be favorable to analysts who are not as familiar with KGs, or a KG's data. P4 echoed this sentiment:

\begin{displayquote}
    \textit{It's possible the user doesn't know anything about the knowledge graph. The LLM could generate at least some basic knowledge about the data or give a summary for the KG. What are the different properties, attributes, structures, relationships, entities? Of course, this might be hard if the KG is very large.} -P4
\end{displayquote}

Other examples of recommendations that an LLM could support, based our discussion with P4 and P5, included: ``\textit{Recommend everything that could be possibly related to the data I care about,}'' ``\textit{Suggest 15 more examples of data that is formatted in this specific way,}'' ``\textit{Suggest a query for me based on this useful domain knowledge,}'' and ``\textit{Tell me other queries I should be running based on what I've already done.}''

\section{Preferred Modalities for LLM Assistants}
\label{sec:design-considerations}
\begin{figure}[h]
    \centering
    \includegraphics[width=0.99\linewidth]{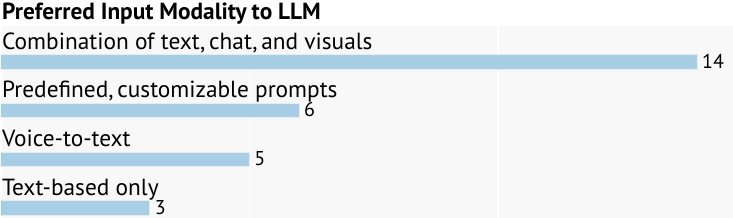}
    \hfill 
    \medbreak
    
    \includegraphics[width=0.99\linewidth]{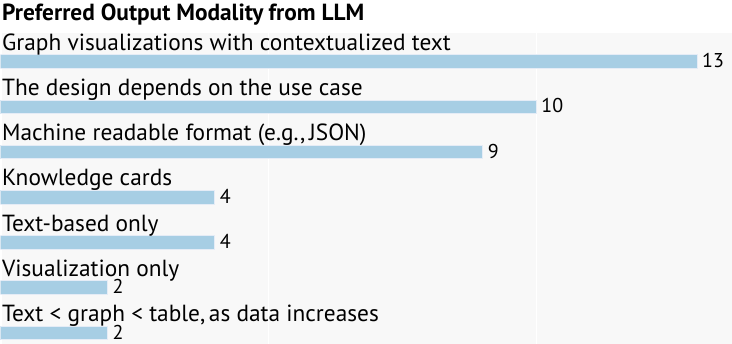}
    \caption{Responses from participants on how they would prefer to interface with the LLM (top) and how they would prefer the LLM to return its results when interfacing with the KG (bottom). We discuss each of these modalities in Section~\ref{sec:design-considerations}.}
    \label{fig:q3}
\end{figure}

We asked participants to specify their preferred methods for interacting or communicating with an LLM during their KG workflows, as well as their preferences concerning how it should display its outputs from the KG. Figure~\ref{fig:q3} shows the response breakdown.


    

\subsection{Interaction \& Inputs to the LLM}
\label{sec:input-modality}

Our second question elicited feedback from participants on the input modalities that would best fit into their own analysis workflows. In general, participants most frequently requested a combination of both text and visualization: ``\textit{Text feels natural because ChatGPT is the big name currently, but I think visual components make a lot of sense for the KG application. I lean towards a hybrid approach}'' (P10). We outline feedback for each modality. 

\subsubsection{Guiding \& Prompting LLMs through Visual Interactions}

Overwhelmingly, participants were most excited about interacting with an LLM through a combination of text and visual display (14/20). Typically this combination was described as a chat-based widget built on top of the applications they frequently used for KG exploration and analysis. For example, P14 told us ``\textit{having a chatbox within Gephi would be ideal,}'' where the visuals (or graph visualizations) help contextualize the user's questions.  

Participants also requested the ability to automatically generate prompts or questions for the LLM by interacting with the visualization directly. A common interaction discussed was direct manipulation of graphical elements to indicate a request for more information about that data point: ``\textit{If I click part of the graph, related text should become more prominent}'' (P1). Similarly, P11 suggested that ``\textit{if I circle [lasso] points on a map, the LLM should be able to filter and summarize query results about them.}''

P11 recommended that ``\textit{global questions}'' (or those exploratory in nature) could be asked in a chat interface, and visualizations could be supplied to help refine those questions into well-specified ones. 

\subsubsection{Prebuilt Prompts and Workflow Templates}
\label{sec:workflow-templates}

Six participants mentioned that rather than engaging with the LLM from a blank chatbox, they wanted to choose from predefined workflows or ``\textit{widgets that encapsulate prompts and parameters}'' (P1). These predefined workflows can be tailored to the user's analysis task, for example, to ``\textit{build custom insights with guided prompts}'' (P9). 

Even participants with prompt engineering expertise noted this feature as desirable for both novices and experts:

\begin{displayquote}
    \textit{A standalone chatbox works, but prompting is a pain. It would be nice to have prebuilt prompts or buttons to start off with.} -P14
\end{displayquote}

Three of our participants specifically described these predefined workflow templates (or prebuilt prompts) as individual \textit{buttons} on a dashboard that ``\textit{represent common KG tasks}'' (P8).

As described in interviews, clicking on a button like ``\textit{Generate Insights}'' would pre-populate a prompt in the user's chatbox, which could then be customized based on their needs. 

\subsubsection{Natural Language Questions through Text \& Voice}
Eight total participants wanted to use natural language to interact with the LLM, with 3/20 preferring to use written text exclusively. P8 described a text-based chat interface as ``\textit{the most natural method to interact with the LLM,}'' and noted that LLMs are one of the few methods to extract structure from free-form text successfully.

Surprisingly, more participants (5/20) wanted the ability to use voice-to-text to interact with the LLM. P20 suggested that a ``\textit{speech recognition system}'' could be used to interpret the user's questions to the LLM (similar to NLIs that support speech-to-text\cite{cox2001multi, setlur2016eviza}). Frequently when we asked participants to specify when they would prefer to use voice instead of text and vice versa, they told us that it depended on the context and specificity of their desired analysis:

\begin{displayquote}
    \textit{Text is best for complicated, precise, thoughtful questions. You can think through the sentence you want to use. But for everyday users you want to use speech, like for buying something online.} -P20
\end{displayquote}

As discussed in Section~\ref{sec:qa-task}, users want an LLM to help them ask well-defined \textit{and} open-ended questions to the KG. If the user is in an exploratory phase of analysis, where iterative refinement is necessary before queries can be executed to the KG, then voice-to-text might be the preferred modality when communicating with the LLM. In contrast, users with specific questions may prefer to use written text.

\subsection{Results Presentation and Communication}
\label{sec:output-modality}

For our third question, we asked participants how they would like the LLM to output its results. 
Primarily, participants wanted a hybrid dashboard in which the results of a KG query are visualized as a graph or subgraph, and the LLM summarizes the results in natural language text to contextualize the KG's outputs. 

Half of our participants emphasized that the output modality 
would heavily depend on the task they are trying to accomplish, ``\textit{if I need [the LLM] to construct queries for me, a text-based format is good. If I am looking for a certain part of data, then a graph visualization is good}'' (P18), and that ``\textit{there should be an option to select the format, as it depends on the use case}'' (P16).

\subsubsection{Text, Graphs, and Tables}

Ten participants told us that whether the LLM outputs its results as text, as graph visualizations, or as tables depends on how much data is being communicated:

\begin{displayquote}
    \textit{It depends on the amount of data being viewed. For a small amount of data (a few nodes and links), text is probably easiest. For a medium amount of data (several nodes, many links) graph-based visualizations become easier to interpret. And when the amount of data exceeds what can be easily visualized, tabular formats are the only option left.} -P8
\end{displayquote}

This overwhelming feedback suggests that there might be an ideal threshold for which modality an LLM (or even a KG) displays its results, depending on the size of the data returned from the query.

Every participant who wanted a combination of text and visualization described a two-panel design in which one panel was used to interface with the LLM, and the other panel was used to show visualizations of the KG data that was being referenced by the LLM.

P2 and P3 told us that relevant ``\textit{portions of the LLM text should be highlighted in the same color as highlighted portions of the knowledge graph.}'' We used this feedback to help design the interface illustrated in Figure~\ref{fig:llm-justification-wireframe}. P2 also suggested that clicking on portions of LLM text should zoom in on the same data on a subgraph, accompanied by ``\textit{contextualizing, prominent annotations.}''

Participants also told us that an LLM outputting query results as visualizations could speed up their analyses: ``\textit{The biggest challenge for the KG tools that I have used is being able to quickly create visuals of the data}'' (P14). Visual outputs could also help with sanity-checking the LLM's response, and help build confidence in the query results: ``\textit{Visualization is a gut check\ldots it gives us confidence. You don't want just the query results. You also want some context to say whether it makes sense, or is familiar to us}'' (P14). 

Two participants told us they \textit{only} wanted to see visual KG data outputted from the LLM. P14 suggested that text alone would be insufficient for interpreting the LLM's outputs: ``\textit{Visualizations are nice\ldots especially since they seem to convey an idea better than text alone. I would definitely prefer a visualization over just text.}''

Finally, visualizations were recommended for helping to show ``\textit{related topics or semantically similar entities and relationships}'' (P7). P14 told us that since their data science team works in a domain that they are not subject matter experts in, being able to see visualizations of neighboring data in the KG helps guide their analyses: ``\textit{there are some genes that we're very comfortable with\ldots if we see certain genes around them that we're also familiar with, then we know we're going down the right track.}'' 

\subsubsection{Executive Summaries}

Executive summaries were requested to help quickly explain the data retrieved or asked about in the KG. Similar to the task \textit{Make Graphics for Presentations and Report} (Section~\ref{sec:task-make-graphics}), participants wanted the LLM to output its results in a format that they could easily present to stakeholders. 

Four of our participants also expressed interest in summary sheets or \textit{knowledge cards} -- high-level visual representations of KG data\cite{li2024kgs} -- that can help users quickly understand the most important details of their query results:

\begin{displayquote}
    \textit{Maybe generating summary sheets would be more comprehensible than subgraphs. I'm imagining seeing something as a Wikipedia-like page generated from a KG as opposed to working with the KG directly.} -P1
\end{displayquote}

While high-level summaries and visual representations were requested, participants also noted that comprehensive details from the LLM would be necessary when ``\textit{churning numbers}'' or performing complex modeling tasks. In other words, receiving summary outputs was often desired for the sake of presentation and rarely for actual downstream analyses.

\subsubsection{Machine Readable Formats}

Participants who use KGs for modeling purposes tended to want the LLM to output results in a machine-readable format, e.g., tabular-based CSVs or JSON, ``\textit{suitable for further processing and analysis}'' (P10). P7 told us specifically that ``\textit{because of my analysis role, I'm primarily interested in machine-readable JSON data}.'' Nine of our participants wanted the LLM's output to be machine-readable. 

That said, similar to the preference of showing tables over graph visualizations when the data exceed a certain number of nodes and edges, participants told us they would want ``tabular-based outputs like CSVs if the LLM is retrieving large amounts of data'' (P18). P5 told us, ``\textit{a machine readable format would be good if my task is to primarily interface with models instead of people.}'' 

\section{Potential Pitfalls \& Roadblocks}
\label{sec:weaknesses}

\begin{figure}[h]
    \centering
    \includegraphics[width=0.99\linewidth]{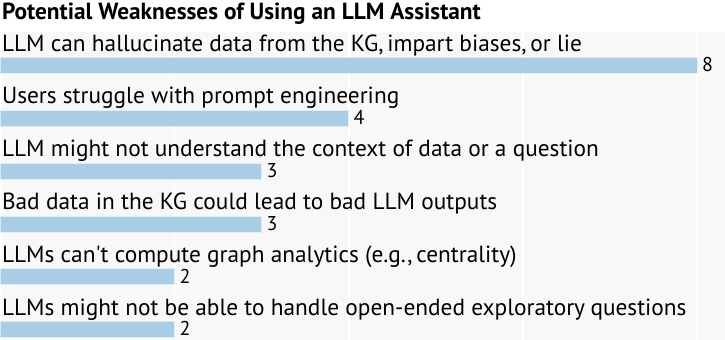}
    \caption{Results of our participants' responses on what weaknesses an LLM might hold when helping users interface with a knowledge graph. 
    We discuss further in Section~\ref{sec:weaknesses}.}
    \label{fig:q2}
\end{figure}

We discuss the potential pitfalls and roadblocks of using LLMs as assistants for KG exploratory analysis. Figure~\ref{fig:q2} shows the code breakdown. Some of these challenges are inherent to the use of LLMs altogether. In other cases, our study participants either anticipated or have already directly experienced the challenges of using LLMs for KG exploration in their own work. We detail these to better inform the future design of systems that integrate LLMs for KGs.

\subsection{Inherent Weaknesses}
\label{sec:llm-weaknesses}
We briefly outline the weaknesses identified by participants below, but point to the many survey papers on this topic (e.g.,\cite{liu2023trustworthy, hadi2023survey}) for more technical details.

\smallbreak 
\noindent 
\textbf{False Information:} Participants were primarily concerned about hallucinations, biases, and ``\textit{intentional lies}'' from LLMs (8/20):
\begin{quote}
    \textit{In addition to hallucinations (and possibly an extension of hallucinations), LLMs can intentionally lie if there is some perceived incentive to do so. Beyond deception, LLMs can produce biased information. 
    Artificial intelligence has long struggled with racist and sexist bias. LLMs are no different.} -P7
\end{quote}

These issues can be overtly troublesome if the user's domain works directly with humans, e.g., social science, as LLMs are known to perpetuate racial prejudices and make problematic judgements\cite{hofmann2024dialect}. Participants also warned us of the flip side of this problem -- using an LLM to confirm your own biases, or outright distrust its outputs if it does not align with your own beliefs: ``\textit{A lot of them [users] look for results to confirm their hypotheses. If the results don't match up with their priors, then they don't trust the results}'' (P13).

Emergent work in hallucination mitigation\cite{wen2023mindmap, li2024enhancing} may partially address these issues; however, they remain an unresolved area of research.

\smallbreak 
\noindent 
\textbf{Prompt Engineering:}
Depending on the model and task complexity, prompting LLMs can be brittle, drastically impact LLM outputs, and require step-by-step chain-of-thought guidance\cite{arora2023ask, chainOfThought}. Four of our participants were concerned about 
difficulties with LLM prompt engineering, and 
spending considerable time crafting the \textit{perfect} prompt:

\begin{quote}
    \textit{There's a huge bias in prompting, and I want to make sure I'm not leaving a better result on the table because I didn't prompt the LLM in the best way.} -P14
\end{quote}

This challenge could be addressed in part by creating predefined prompt templates that have already been tested to perform well on their associated analysis task, as discussed in Section~\ref{sec:workflow-templates}.

\smallbreak 
\noindent 
\textbf{Maintaining User's Semantic or Domain-Specific Intent:} Three participants were concerned with the LLM's ability to understand the context of their question or data. These participants told us that their use case is very domain-specific, and an LLM may overly generalize their intended task. Two participants thought an LLM may be unable to scope down open-ended questions into KG queries, incurring information loss. As potential preventative measures, we suggest enforcing conversational agents to participate in back-and-forth dialogue to uncover the user's intent (Section~\ref{sec:task-refine}), or implementing Retrieval Augmented Generation (RAG) to improve domain-specific question-answering\cite{li2024enhancing}. Arefeed et al. describe a cost-efficient solution\cite{arefeen2024leancontext}.

\smallbreak 
\noindent 
\textbf{Poor KG Data Quality:} Although not an inherent weakness of LLMs, three participants were concerned that poor quality KG data would inevitably lead to poor results from the LLM. P19 said, \textit{``there is often a lot of junk in KGs, and I don't know if an LLM would be able to separate the wheat from the chaff}.'' Issues with KG data have been well-documented\cite{li2024kgs} and can be anticipated in future system designs. 

\subsection{Mismatched Expectations for an LLM}
Prior research has shown that people's mental model of an LLM (or conversational agent) can be extremely unrealistic -- with some users of ChatGPT describing it as ``\textit{kinds of magic that I don't know}''\cite{zhang2023s}. Our participants suggested that interfaces using an LLM to automate tasks should caution users of its known weaknesses:

\begin{displayquote}
    \textit{I've previously used an LLM to generate descriptions of common household items\ldots 
    the mistakes that the LLM made were clustered on specific categories. It made pretty egregious statements regarding which genders could be found in which rooms. So I think having a good understanding of where the LLM struggles and providing a warning up front to users would be valuable.} -P8
\end{displayquote}

P17 told us labeling LLMs as assistants altogether can further mislead users on the competencies of an LLM:

\begin{displayquote}
    \textit{I don't believe LLMs should be personified as ``assistants'' at all. The ELIZA effect is strong and can lead end users to think that the LLM ``knows'' the information in a real-world sense beyond the facts encoded in the KG, which isn't true and can be misleading and potentially harmful.} 
    -P17
\end{displayquote}

Since natural language interfaces are typically designed for non-experts\cite{li2014constructing}, it is essential that \textit{all} future LLM-driven visual analysis systems are transparent in what the LLM is and is not capable of.  The potential pitfalls outlined here serve as a starting point concerning which weaknesses should be addressed in the design of these interfaces.

\section{Design Considerations \& Opportunities}
\label{sec:roadmap}

We outline promising directions for the integration of LLMs as assistants to KG exploration and analysis.

\subsection{LLM and KG Workflows}
\label{sec:workflow}

Exactly \textit{how} an LLM interfaces with the user and KG was a frequent topic discussed in our interviews. Figure~\ref{fig:workflow} illustrates a modular workflow derived from the findings of our study, where the LLM acts as an interpreter for the user's stated task, and coordinates the execution of potentially many functions to the KG.  We refer to these collections of functions as \textit{task templates}, where a template is designed to contain potentially several step-by-step instructions or algorithms for the LLM to run, based on one overarching analytic goal or user task. 

For example, take the exemplar use case illustrated in Figure~\ref{fig:llm-justification-wireframe}. Using our workflow in Figure~\ref{fig:workflow}, the LLM would first interpret the user's intended task from their initial statement: ``\textit{What are the connections between Source Node and Target Node?}'' The LLM then selects the task template that best matches the goal ``\textit{identify all paths between entities.}'' Using that task template, the LLM would follow a set of system-defined instructions, such as:


\begin{enumerate}[topsep=2pt, partopsep=0pt,itemsep=1pt,parsep=2pt]
    \small
    \item \texttt{Extract the entity names the user wants to identify paths between (i.e. \textit{Source Node} and \textit{Target Node})}
    \item \texttt{Find their IDs associated in the KG}
    \item \texttt{Call the system-instructed NetworkX functions, or execute a Cypher query, to find paths between the nodes}
    \item \texttt{Filter the results into a more digestible subgraph, if the results exceed the system-specified threshold}
    \item \texttt{Pass the subgraph results to a graph visualizer}
    \item \texttt{Respond to the user in chat and summarize the results}
\end{enumerate}

Importantly, the simple instruction of ``\textit{write me a Cypher query}'' could produce poor results, as LLMs may not be trained to produce such structured text:

\begin{displayquote}
   ``\textit{Conversational LLMs are really good at semantic parsing, writing coherent text\ldots But when you ask it to write a query, that's not something it's naturally good at. It doesn't have enough training data on that, it's good at token by token, etc. The LLM is not good at generating structured information.}'' -P20
\end{displayquote}

LLMs that are fine-tuned for the task of writing queries (e.g.,\cite{yang2023llm}) can overcome this specific challenge. However, it may be computationally infeasible to fine-tune an LLM such that it is performant on \textit{all} KG tasks. For example, it may not be practical to try to train an LLM to memorize all the entity and relation IDs in a KG.
Therefore, we recommend that developers of LLM-driven KG systems consider workflows in which the LLM can truly take on the role of an \textit{assistant} or \textit{coordinator}. Consider what the LLM is inherently good at -- interpreting text, refining questions, producing summaries -- and allow it to take on those tasks in the user's workflow. For all other tasks that an LLM might struggle with, explore how it can pass the baton across different parts of the system to help achieve the user's goals. 


\subsection{LLM and KG Systems} 
\label{sec:roadmap-systems}

The tasks and interface modalities outlined in Sections~\ref{sec:tasks} and~\ref{sec:design-considerations} help to broadly inform the design of future LLM and KG systems. Here, we discuss opportunities for different goal-oriented systems that participants frequently brought up in our interviews. 

\smallbreak 
\noindent 
\textbf{Data Coordination for AI/ML:} 
Poor data quality is consistently cited as the major reason why AI/ML collaborations (as well as AI/ML models altogether) fail\cite{hong2020human, sambasivan2021everyone} -- this challenge does not exclude KG usage\cite{li2024kgs}. P14 told us that one way his data science team has overcome this challenge is by using an LLM to search through different data catalogs, databases, and ``\textit{pull that disparate data into a knowledge graph.}'' In this case, a system could be used to coordinate data across multiple sources to keep a KG consistently refreshed for AI/ML tasks.

\smallbreak 
\noindent 
\textbf{Intelligent Query Builders:} A common criticism of NLIs contributed in research is their lack of ``intelligence''\cite{setlur2016eviza, dhamdhere2017analyza, karanikolas2023large}. Users wanted an LLM-driven NLI to ask them follow-up questions, provide recommendations for relevant data, and extract meaningful context about the data domain. Our findings underpin how essential support is for these question-answering and data retrieval tasks (Section~\ref{sec:qa-task}). 
In this case, a system could be responsible for helping extract questions from a user's analytic goals, including the refinement of open-ended questions into well-formed KG queries.

\smallbreak 
\noindent 
\textbf{Text Oracle:} Participants wanted a system in which an LLM is responsible for finding, summarizing, and providing explanations with text. From their own experience, LLMs had shown promise in extracting potential KG features from research articles and providing executive summaries from text documents (Section~\ref{sec:task-improve}). However, because an LLM may produce false information (Section~\ref{sec:weaknesses}), such as nonexistent citations or sources, participants had difficulty using an LLM to explain itself or its outputs. A holistic system that combines the explainability features of KGs\cite{lecue2020role, tiddi2020knowledge} could potentially address these issues while leveraging the text extraction capabilities of an LLM.

\smallbreak 
\noindent 
\textbf{Analysis Tracker:} 
Participants asked for a system where the LLM was responsible for tracking KG changes to inform the user's downstream analyses. This can be helpful if the KG is a collaborative effort, such that many builders are maintaining its infrastructure. By having an LLM enabled to track these changes, a KG builder could ask questions like, ``\textit{how has the KG changed in the past 3 days? What new data has been added or deleted?}'' For analysts, changes in the KG may also change the results of their prior analyses. An LLM and KG system could be used to alert data scientists how their previous findings may be out-of-date, or could automatically recalculate metrics given changes in the KG -- alleviating this burden for data scientists altogether.

\smallbreak 
\noindent 
\textbf{Content Generation:} P2 told us he represents his \textit{Dungeons \& Dragons} campaigns in a knowledge graph format, with written text to describe the worlds he creates. He wanted an LLM-driven system to be able to read those texts, have access to his graph-based campaigns, and provide him with new world-building capabilities, e.g., suggest stories, characters, and missions for his games. 
We note that using an LLM system in this way is likely closer to supporting Brehmer and Munzner's \textit{Enjoy} task\cite{brehmer2013multi}, suggesting that users do not want an LLM to strictly help them with data analysis. Similar work has been done to use LLMs to create content for role-playing and educational games\cite{wu2023autogen}.

\smallbreak 
\noindent 
\textbf{Spur New Ideas:} Although LLMs were criticized for confirming users' biases (Section~\ref{sec:llm-weaknesses}), four participants talked about the benefits of using an LLM-driven system to help spur new ideas for their analyses: ``\textit{LLMs tend to be creative. Anytime you want to retrieve information [from it], you combine parameters to generate an answer. The generative aspect of the LLM is creative!}'' Furthermore, participants remarked that an LLM can help challenge existing biases, particularly the prior beliefs that domain experts hold: ``\textit{An LLM is nice because it adds some creativity on top of what the lab scientists [SMEs] think about. The scientists usually look at one kind of connection, but the LLMs can show them new kinds of connections}'' (P13).

\section{Discussion}
\label{sec:discussion}

The purpose of our study was to distill the tasks, interaction capabilities, communication modalities, and analysis workflows an LLM can best support for KG users. Each of these components serve as the building blocks of our proposed roadmap, and can be leveraged as a starting point for the design of NLIs, IUIs, and visual analytics systems that integrate an LLM assistant. That said, further research is needed to concretize our preliminary roadmap. 


For example, future work should examine whether our proposed roadmap applies more broadly to structured knowledge sources beyond KGs. An LLM might suffer from additional (or fewer) limitations when querying a data source that is perhaps less well-structured than a knowledge graph. 
By studying how LLMs can lower the barrier of data management, there is potential to overcome a number of collaboration challenges in human-AI workflows\cite{chiang2023two} and analysis models\cite{pirolli2005sensemaking, tukey1977exploratory}. Relatedly, recent work in data management\cite{fernandez2023large} and relational databases\cite{li2023can} discuss the potential of LLMs to serve as the `interface' between a user and large-scale enterprise databases. 

Our roadmap focuses on using LLMs as assistants to KGs; however, the dual use of an LLM and KG can serve multiple complementary purposes. Beyond helping to ground an LLM's outputs using data from the KG (discussed in Section~\ref{sec:llm-justification}), a KG can also be used to fact-check the LLM or provide more up-to-date results than what the LLM was previously trained on. Research in LLM development\cite{2024_unifying_llms_and_kgs} examines the benefits of introducing a KG into an LLM's architecture. Future work can examine extending our roadmap to include the other direction of our research: using KGs to assist LLM-driven interfaces.

Finally, our preliminary roadmap does not include recommendations for the evaluation of LLM-driven KG systems. Prior work in NLIs\cite{srinivasan2021snowy} and LLMs in HCI\cite{rapp2023collaborating} can begin to illuminate this research direction, but future work will need to understand how to best test the efficacy of these systems. Moreover, our roadmap does not include recommendations for handling the safety risks or ethical concerns of using LLMs\cite{zhang2023s}, although these critical topics demand further investigation. Along both of these lines of work, future research should address the `best ways' for an LLM to explain itself\cite{danry2023dont} when returning results from its interactions with the KG. We hope that the findings of our study, as well as the design considerations outlined in our paper, can serve as a springboard to all of these emergent research opportunities.

\section{Conclusion}

This paper presented findings from a mixed methods study with 20 KG and LLM experts to inform a preliminary roadmap for the design of KG visual data analysis systems with LLM assistants. We distilled the most common tasks that KG and LLM experts desired assistance with, as well as preferred input and output modalities when interacting and communicating with an LLM. 
Our results show that users want an LLM to facilitate data analytic tasks such as question-answering, improving data quality, summarizing query results, and curating custom datasets from a KG. Participants noted that LLMs excel in many aspects of these tasks, such as conducting back-and-forth dialog to refine analysis goals, and can be used to overcome frequent KG challenges. In general, we found that our participants prefer a mixed modality approach to communicating with and receiving results from the LLM. Many participants wanted pre-built, customizable prompts that were tailored to specific analysis tasks, while others wanted the ability to prompt the LLM with their interactions with a visualization. From the analysis of our results, we outlined future opportunities in this space, including collaborative LLM and KG workflows and system design considerations.

\acknowledgments{%
We sincerely thank each of the participants who took the time to support our study.
\medbreak
\noindent 
DISTRIBUTION STATEMENT A. Approved for public release. Distribution is unlimited.
This material is based upon work supported by the Combatant Commands under Air Force Contract No. FA8702-15-D-0001. Any opinions, findings, conclusions or recommendations expressed in this material are those of the author(s) and do not necessarily reflect the views of the Combatant Commands. © 2024 Massachusetts Institute of Technology. Delivered to the U.S. Government with Unlimited Rights, as defined in DFARS Part 252.227-7013 or 7014 (Feb 2014). Notwithstanding any copyright notice, U.S. Government rights in this work are defined by DFARS 252.227-7013 or DFARS 252.227-7014 as detailed above. Use of this work other than as specifically authorized by the U.S. Government may violate any copyrights that exist in this work.
}

\bibliographystyle{abbrv-doi-hyperref}

\bibliography{main}

\begin{thebibliography}{10}

\bibitem{abu2021domain}
B.~Abu-Salih.
\newblock Domain-specific knowledge graphs: A survey.
\newblock {\em J. Netw. Comput. Appl.}, 185:103076, 2021. \href{https://doi.org/10.1016/j.jnca.2021.103076}
{doi: {{%
10\hspace{.1pt}\discretionary{.}{%
}{.}\hspace{.4pt}1016\discretionary{/}{%
}{/}j\hspace{.1pt}\discretionary{.}{%
}{.}\hspace{.4pt}jnca\hspace{.1pt}\discretionary{.}{%
}{.}\hspace{.4pt}2021\hspace{.1pt}\discretionary{.}{%
}{.}\hspace{.4pt}103076}}}


\bibitem{agrawal2023can}
G.~Agrawal, T.~Kumarage, Z.~Alghami, and H.~Liu.
\newblock Can knowledge graphs reduce hallucinations in llms?: A survey.
\newblock {\em arXiv preprint arXiv:2311.07914}, 2023.

\bibitem{akbaba2023two}
D.~Akbaba and M.~Meyer.
\newblock “two heads are better than one”: Pair-interviews for visualization.
\newblock In {\em 2023 IEEE Visualization and Visual Analytics (VIS)}, pp. 206--210. IEEE, 2023.

\bibitem{alkhamissi2022review}
B.~AlKhamissi, M.~Li, A.~Celikyilmaz, M.~Diab, and M.~Ghazvininejad.
\newblock A review on language models as knowledge bases.
\newblock {\em arXiv preprint arXiv:2204.06031}, 2022. \href{https://doi.org/10.48550/arXiv.2204.06031}
{doi: {{%
10\hspace{.1pt}\discretionary{.}{%
}{.}\hspace{.4pt}48550\discretionary{/}{%
}{/}arXiv\hspace{.1pt}\discretionary{.}{%
}{.}\hspace{.4pt}2204\hspace{.1pt}\discretionary{.}{%
}{.}\hspace{.4pt}06031}}}


\bibitem{arefeen2024leancontext}
M.~A. Arefeen, B.~Debnath, and S.~Chakradhar.
\newblock Leancontext: Cost-efficient domain-specific question answering using llms.
\newblock {\em Natural Language Processing Journal}, p. 100065, 2024.

\bibitem{arora2023ask}
S.~Arora, A.~Narayan, M.~F. Chen, L.~Orr, N.~Guha, K.~Bhatia, I.~Chami, and C.~Re.
\newblock Ask me anything: A simple strategy for prompting language models.
\newblock In {\em The Eleventh International Conference on Learning Representations}, 2023.

\bibitem{Aurisano2016Articulate2}
J.~Aurisano, A.~Kumar, A.~Gonzales, J.~Leigh, B.~DiEugenio, and A.~Johnson.
\newblock Articulate 2 : Toward a conversational interface for visual data exploration.
\newblock In {\em Proc. VIS}, 2016.

\bibitem{berners2001semantic}
T.~Berners-Lee, J.~Hendler, and O.~Lassila.
\newblock The semantic web.
\newblock {\em Sci. Am.}, 284(5):34--43, 2001.

\bibitem{Bonatti:2019:Knowledge}
P.~A. Bonatti, S.~Decker, A.~Polleres, and V.~Presutti.
\newblock {Knowledge Graphs: New Directions for Knowledge Representation on the Semantic Web (Dagstuhl Seminar 18371)}.
\newblock {\em Dagstuhl Reports}, 8(9):29--111, 2019. \href{https://doi.org/10.4230/DagRep.8.9.29}
{doi: {{%
10\hspace{.1pt}\discretionary{.}{%
}{.}\hspace{.4pt}4230\discretionary{/}{%
}{/}DagRep\hspace{.1pt}\discretionary{.}{%
}{.}\hspace{.4pt}8\hspace{.1pt}\discretionary{.}{%
}{.}\hspace{.4pt}9\hspace{.1pt}\discretionary{.}{%
}{.}\hspace{.4pt}29}}}


\bibitem{braun2006using}
V.~Braun and V.~Clarke.
\newblock Using thematic analysis in psychology.
\newblock {\em Qual. Res. Psychol.}, 3(2):77--101, 2006. \href{https://doi.org/10.1191/1478088706qp063oa}
{doi: {{%
10\hspace{.1pt}\discretionary{.}{%
}{.}\hspace{.4pt}1191\discretionary{/}{%
}{/}1478088706qp063oa}}}


\bibitem{brehmer2013multi}
M.~Brehmer and T.~Munzner.
\newblock A multi-level typology of abstract visualization tasks.
\newblock {\em IEEE transactions on visualization and computer graphics}, 19(12):2376--2385, 2013.

\bibitem{brown2020language}
T.~Brown, B.~Mann, N.~Ryder, M.~Subbiah, J.~D. Kaplan, P.~Dhariwal, A.~Neelakantan, P.~Shyam, G.~Sastry, A.~Askell, S.~Agarwal, A.~Herbert-Voss, G.~Krueger, T.~Henighan, R.~Child, A.~Ramesh, D.~Ziegler, J.~Wu, C.~Winter, C.~Hesse, M.~Chen, E.~Sigler, M.~Litwin, S.~Gray, B.~Chess, J.~Clark, C.~Berner, S.~McCandlish, A.~Radford, I.~Sutskever, and D.~Amodei.
\newblock Language models are few-shot learners.
\newblock In {\em Adv. Neural Inf.}, vol.~33, pp. 1877--1901. Curran Associates, Inc., 2020.

\bibitem{Cashman:2020:CAVA}
D.~{Cashman}, S.~{Xu}, S.~{Das}, F.~{Heimerl}, C.~{Liu}, S.~R. {Humayoun}, M.~{Gleicher}, A.~{Endert}, and R.~{Chang}.
\newblock Cava: A visual analytics system for exploratory columnar data augmentation using knowledge graphs.
\newblock {\em IEEE Trans. Vis. Comput. Graph.}, 27(2):1731--1741, 2021. \href{https://doi.org/10.1109/TVCG.2020.3030443}
{doi: {{%
10\hspace{.1pt}\discretionary{.}{%
}{.}\hspace{.4pt}1109\discretionary{/}{%
}{/}TVCG\hspace{.1pt}\discretionary{.}{%
}{.}\hspace{.4pt}2020\hspace{.1pt}\discretionary{.}{%
}{.}\hspace{.4pt}3030443}}}


\bibitem{chen2020knowledge}
Z.~Chen, Y.~Wang, B.~Zhao, J.~Cheng, X.~Zhao, and Z.~Duan.
\newblock Knowledge graph completion: A review.
\newblock {\em Ieee Access}, 8:192435--192456, 2020.

\bibitem{chen2023beyond}
Z.~Chen, C.~Zhang, Q.~Wang, J.~Troidl, S.~Warchol, J.~Beyer, N.~Gehlenborg, and H.~Pfister.
\newblock Beyond generating code: Evaluating gpt on a data visualization course.
\newblock In {\em 2023 IEEE VIS Workshop on Visualization Education, Literacy, and Activities (EduVis)}, pp. 16--21, 2023. \href{https://doi.org/10.1109/EduVis60792.2023.00009}
{doi: {{%
10\hspace{.1pt}\discretionary{.}{%
}{.}\hspace{.4pt}1109\discretionary{/}{%
}{/}EduVis60792\hspace{.1pt}\discretionary{.}{%
}{.}\hspace{.4pt}2023\hspace{.1pt}\discretionary{.}{%
}{.}\hspace{.4pt}00009}}}


\bibitem{chiang2023two}
C.-W. Chiang, Z.~Lu, Z.~Li, and M.~Yin.
\newblock Are two heads better than one in ai-assisted decision making? comparing the behavior and performance of groups and individuals in human-ai collaborative recidivism risk assessment.
\newblock In {\em In Proc. CHI 2023}, CHI '23. Association for Computing Machinery, New York, NY, USA, 2023. \href{https://doi.org/10.1145/3544548.3581015}
{doi: {{%
10\hspace{.1pt}\discretionary{.}{%
}{.}\hspace{.4pt}1145\discretionary{/}{%
}{/}3544548\hspace{.1pt}\discretionary{.}{%
}{.}\hspace{.4pt}3581015}}}


\bibitem{paulheim2017knowledge}
P.~Cimiano and H.~Paulheim.
\newblock Knowledge graph refinement: A survey of approaches and evaluation methods.
\newblock {\em Semant. Web}, 8(3):489–508, 2017. \href{https://doi.org/10.3233/SW-160218}
{doi: {{%
10\hspace{.1pt}\discretionary{.}{%
}{.}\hspace{.4pt}3233\discretionary{/}{%
}{/}SW\discretionary{%
}{-}{-}160218}}}


\bibitem{cox2001multi}
K.~Cox, R.~E. Grinter, S.~L. Hibino, L.~J. Jagadeesan, and D.~Mantilla.
\newblock A multi-modal natural language interface to an information visualization environment.
\newblock {\em International Journal of Speech Technology}, 4:297--314, 2001.

\bibitem{curry2009qualitative}
L.~A. Curry, I.~M. Nembhard, and E.~H. Bradley.
\newblock Qualitative and mixed methods provide unique contributions to outcomes research.
\newblock {\em Circulation}, 119(10):1442--1452, 2009.

\bibitem{danry2023dont}
V.~Danry, P.~Pataranutaporn, Y.~Mao, and P.~Maes.
\newblock Don’t just tell me, ask me: Ai systems that intelligently frame explanations as questions improve human logical discernment accuracy over causal ai explanations.
\newblock In {\em In Proc. CHI 2023}, CHI '23. Association for Computing Machinery, New York, NY, USA, 2023. \href{https://doi.org/10.1145/3544548.3580672}
{doi: {{%
10\hspace{.1pt}\discretionary{.}{%
}{.}\hspace{.4pt}1145\discretionary{/}{%
}{/}3544548\hspace{.1pt}\discretionary{.}{%
}{.}\hspace{.4pt}3580672}}}


\bibitem{decuir2011developing}
J.~T. DeCuir-Gunby, P.~L. Marshall, and A.~W. McCulloch.
\newblock Developing and using a codebook for the analysis of interview data: An example from a professional development research project.
\newblock {\em Field Methods}, 23(2):136--155, 2011. \href{https://doi.org/10.1177/1525822X10388468}
{doi: {{%
10\hspace{.1pt}\discretionary{.}{%
}{.}\hspace{.4pt}1177\discretionary{/}{%
}{/}1525822X10388468}}}


\bibitem{dhamdhere2017analyza}
K.~Dhamdhere, K.~S. McCurley, R.~Nahmias, M.~Sundararajan, and Q.~Yan.
\newblock Analyza: Exploring data with conversation.
\newblock In {\em In Proc. ACM IUI}, pp. 493--504, 2017.

\bibitem{ehrlinger2016towards}
L.~Ehrlinger and W.~W{\"o}{\ss}.
\newblock Towards a definition of knowledge graphs.
\newblock {\em Proc. ESWC Posters and Demos Track}, 48(1-4):2, 2016.

\bibitem{fernandez2023large}
R.~C. Fernandez, A.~J. Elmore, M.~J. Franklin, S.~Krishnan, and C.~Tan.
\newblock How large language models will disrupt data management.
\newblock {\em In Proc. VLDB}, 16(11):3302--3309, 2023.

\bibitem{francis2018cypher}
N.~Francis, A.~Green, P.~Guagliardo, L.~Libkin, T.~Lindaaker, V.~Marsault, S.~Plantikow, M.~Rydberg, P.~Selmer, and A.~Taylor.
\newblock Cypher: An evolving query language for property graphs.
\newblock In {\em Proc. SIGMOD}, p. 1433–1445. ACM, New York, 2018. \href{https://doi.org/10.1145/3183713.3190657}
{doi: {{%
10\hspace{.1pt}\discretionary{.}{%
}{.}\hspace{.4pt}1145\discretionary{/}{%
}{/}3183713\hspace{.1pt}\discretionary{.}{%
}{.}\hspace{.4pt}3190657}}}


\bibitem{gal2014uncertain}
A.~Gal.
\newblock Uncertain entity resolution: Re-evaluating entity resolution in the big data era: Tutorial.
\newblock {\em Proc. VLDB Endow.}, 7(13):1711–1712, 2014. \href{https://doi.org/10.14778/2733004.2733068}
{doi: {{%
10\hspace{.1pt}\discretionary{.}{%
}{.}\hspace{.4pt}14778\discretionary{/}{%
}{/}2733004\hspace{.1pt}\discretionary{.}{%
}{.}\hspace{.4pt}2733068}}}


\bibitem{Gan2021NaturalSM}
Y.~Gan, X.~Chen, J.~Xie, M.~Purver, J.~R. Woodward, J.~Drake, and Q.~Zhang.
\newblock Natural {SQL}: Making {SQL} easier to infer from natural language specifications.
\newblock In {\em Proc. EMNLP}, pp. 2030--2042. ACL, Punta Cana, Dominican Republic, 2021. \href{https://doi.org/10.18653/v1/2021.findings-emnlp.174}
{doi: {{%
10\hspace{.1pt}\discretionary{.}{%
}{.}\hspace{.4pt}18653\discretionary{/}{%
}{/}v1\discretionary{/}{%
}{/}2021\hspace{.1pt}\discretionary{.}{%
}{.}\hspace{.4pt}findings\discretionary{%
}{-}{-}emnlp\hspace{.1pt}\discretionary{.}{%
}{.}\hspace{.4pt}174}}}


\bibitem{hadi2023survey}
M.~U. Hadi, R.~Qureshi, A.~Shah, M.~Irfan, A.~Zafar, M.~B. Shaikh, N.~Akhtar, J.~Wu, S.~Mirjalili, et~al.
\newblock A survey on large language models: Applications, challenges, limitations, and practical usage.
\newblock {\em Authorea Preprints}, 2023.

\bibitem{hofmann2024dialect}
V.~Hofmann, P.~R. Kalluri, D.~Jurafsky, and S.~King.
\newblock Dialect prejudice predicts ai decisions about people's character, employability, and criminality.
\newblock {\em arXiv preprint arXiv:2403.00742}, 2024.

\bibitem{hogan2021knowledge}
A.~Hogan, E.~Blomqvist, M.~Cochez, C.~D’amato, G.~D. Melo, C.~Gutierrez, S.~Kirrane, J.~E.~L. Gayo, R.~Navigli, S.~Neumaier, A.-C.~N. Ngomo, A.~Polleres, S.~M. Rashid, A.~Rula, L.~Schmelzeisen, J.~Sequeda, S.~Staab, and A.~Zimmermann.
\newblock Knowledge graphs.
\newblock {\em ACM Comput. Surv.}, 54(4), 2021. \href{https://doi.org/10.1145/3447772}
{doi: {{%
10\hspace{.1pt}\discretionary{.}{%
}{.}\hspace{.4pt}1145\discretionary{/}{%
}{/}3447772}}}


\bibitem{hong2020human}
S.~R. Hong, J.~Hullman, and E.~Bertini.
\newblock Human factors in model interpretability: Industry practices, challenges, and needs.
\newblock {\em Proc. CHI}, 4(CSCW1), 2020. \href{https://doi.org/10.1145/3392878}
{doi: {{%
10\hspace{.1pt}\discretionary{.}{%
}{.}\hspace{.4pt}1145\discretionary{/}{%
}{/}3392878}}}


\bibitem{huang2023flownl}
J.~Huang, Y.~Xi, J.~Hu, and J.~Tao.
\newblock Flownl: Asking the flow data in natural languages.
\newblock {\em IEEE Trans. Vis. Comput. Graph.}, 29(1):1200--1210, 2023. \href{https://doi.org/10.1109/TVCG.2022.3209453}
{doi: {{%
10\hspace{.1pt}\discretionary{.}{%
}{.}\hspace{.4pt}1109\discretionary{/}{%
}{/}TVCG\hspace{.1pt}\discretionary{.}{%
}{.}\hspace{.4pt}2022\hspace{.1pt}\discretionary{.}{%
}{.}\hspace{.4pt}3209453}}}


\bibitem{karanikolas2023large}
N.~Karanikolas, E.~Manga, N.~Samaridi, E.~Tousidou, and M.~Vassilakopoulos.
\newblock Large language models versus natural language understanding and generation.
\newblock In {\em In Proc. PCI}, pp. 278--290, 2023.

\bibitem{kepuska2018next}
V.~Kepuska and G.~Bohouta.
\newblock Next-generation of virtual personal assistants (microsoft cortana, apple siri, amazon alexa and google home).
\newblock In {\em 2018 IEEE 8th annual computing and communication workshop and conference (CCWC)}, pp. 99--103. IEEE, 2018.

\bibitem{kintzer1977advantages}
F.~C. Kintzer.
\newblock Advantages of open-response questions in survey research.
\newblock {\em Community Junior College Research Quarterly}, 2(1):37--46, 1977.

\bibitem{klein2022bringing}
K.~Klein, J.~F. Sequeda, H.-Y. Wu, and D.~Yan.
\newblock {Bringing Graph Databases and Network Visualization Together (Dagstuhl Seminar 22031)}.
\newblock {\em Dagstuhl Reports}, 12(1):67--82, 2022. \href{https://doi.org/10.4230/DagRep.12.1.67}
{doi: {{%
10\hspace{.1pt}\discretionary{.}{%
}{.}\hspace{.4pt}4230\discretionary{/}{%
}{/}DagRep\hspace{.1pt}\discretionary{.}{%
}{.}\hspace{.4pt}12\hspace{.1pt}\discretionary{.}{%
}{.}\hspace{.4pt}1\hspace{.1pt}\discretionary{.}{%
}{.}\hspace{.4pt}67}}}


\bibitem{koh2024generating}
J.~Y. Koh, D.~Fried, and R.~R. Salakhutdinov.
\newblock Generating images with multimodal language models.
\newblock {\em Advances in Neural Information Processing Systems}, 36, 2024.

\bibitem{lecue2020role}
F.~Lecue.
\newblock On the role of knowledge graphs in explainable {AI}.
\newblock {\em Semant. Web}, 11(1):41--51, 2020. \href{https://doi.org/10.3233/SW-190374}
{doi: {{%
10\hspace{.1pt}\discretionary{.}{%
}{.}\hspace{.4pt}3233\discretionary{/}{%
}{/}SW\discretionary{%
}{-}{-}190374}}}


\bibitem{li2014constructing}
F.~Li and H.~V. Jagadish.
\newblock Constructing an interactive natural language interface for relational databases.
\newblock {\em In Proc. VLDB}, 8(1):73--84, 2014.

\bibitem{li2024kgs}
H.~Li, G.~Appleby, C.~D. Brumar, R.~Chang, and A.~Suh.
\newblock Knowledge graphs in practice: Characterizing their users, challenges, and visualization opportunities.
\newblock {\em IEEE Transactions on Visualization and Computer Graphics}, 30(1):584--594, 2024. \href{https://doi.org/10.1109/TVCG.2023.3326904}
{doi: {{%
10\hspace{.1pt}\discretionary{.}{%
}{.}\hspace{.4pt}1109\discretionary{/}{%
}{/}TVCG\hspace{.1pt}\discretionary{.}{%
}{.}\hspace{.4pt}2023\hspace{.1pt}\discretionary{.}{%
}{.}\hspace{.4pt}3326904}}}


\bibitem{li2023can}
J.~Li, B.~Hui, G.~Qu, B.~Li, J.~Yang, B.~Li, B.~Wang, B.~Qin, R.~Cao, R.~Geng, et~al.
\newblock Can llm already serve as a database interface? a big bench for large-scale database grounded text-to-sqls.
\newblock {\em arXiv preprint arXiv:2305.03111}, 2023.

\bibitem{li2024enhancing}
J.~Li, Y.~Yuan, and Z.~Zhang.
\newblock Enhancing llm factual accuracy with rag to counter hallucinations: A case study on domain-specific queries in private knowledge-bases.
\newblock {\em arXiv preprint arXiv:2403.10446}, 2024.

\bibitem{lin2024inksight}
Y.~Lin, H.~Li, L.~Yang, A.~Wu, and H.~Qu.
\newblock Inksight: Leveraging sketch interaction for documenting chart findings in computational notebooks.
\newblock {\em IEEE Transactions on Visualization and Computer Graphics}, 30(1):944--954, 2024. \href{https://doi.org/10.1109/TVCG.2023.3327170}
{doi: {{%
10\hspace{.1pt}\discretionary{.}{%
}{.}\hspace{.4pt}1109\discretionary{/}{%
}{/}TVCG\hspace{.1pt}\discretionary{.}{%
}{.}\hspace{.4pt}2023\hspace{.1pt}\discretionary{.}{%
}{.}\hspace{.4pt}3327170}}}


\bibitem{lissandrini2022knowledge}
M.~Lissandrini, D.~Mottin, K.~Hose, and T.~B. Pedersen.
\newblock Knowledge graph exploration systems: are we lost?
\newblock In {\em CIDR}, vol.~22, pp. 10--13, 2022.

\bibitem{lissandrini2020graph}
M.~Lissandrini, D.~Mottin, T.~Palpanas, and Y.~Velegrakis.
\newblock Graph-query suggestions for knowledge graph exploration.
\newblock In {\em In Proc. ACM WWW}, pp. 2549--2555, 2020.

\bibitem{liu2023trustworthy}
Y.~Liu, Y.~Yao, J.-F. Ton, X.~Zhang, R.~G.~H. Cheng, Y.~Klochkov, M.~F. Taufiq, and H.~Li.
\newblock Trustworthy llms: a survey and guideline for evaluating large language models' alignment.
\newblock {\em arXiv preprint arXiv:2308.05374}, 2023.

\bibitem{luo2023reasoning}
L.~Luo, Y.-F. Li, G.~Haffari, and S.~Pan.
\newblock Reasoning on graphs: Faithful and interpretable large language model reasoning.
\newblock {\em arXiv preprint arXiv:2310.01061}, 2023.

\bibitem{macqueen1998codebook}
K.~M. MacQueen, E.~McLellan, K.~Kay, and B.~Milstein.
\newblock Codebook development for team-based qualitative analysis.
\newblock {\em CAM j.}, 10(2):31--36, 1998. \href{https://doi.org/10.1177/1525822X980100020301}
{doi: {{%
10\hspace{.1pt}\discretionary{.}{%
}{.}\hspace{.4pt}1177\discretionary{/}{%
}{/}1525822X980100020301}}}


\bibitem{martino2023knowledge}
A.~Martino, M.~Iannelli, and C.~Truong.
\newblock Knowledge injection to counter large language model (llm) hallucination.
\newblock In {\em European Semantic Web Conference}, pp. 182--185. Springer, 2023.

\bibitem{knowledgeInjecitonToCounter}
A.~Martino, M.~Iannelli, and C.~Truong.
\newblock Knowledge injection to counter large language model (llm) hallucination.
\newblock In C.~Pesquita, H.~Skaf-Molli, V.~Efthymiou, S.~Kirrane, A.~Ngonga, D.~Collarana, R.~Cerqueira, M.~Alam, C.~Trojahn, and S.~Hertling, eds., {\em The Semantic Web: ESWC 2023 Satellite Events}, pp. 182--185. Springer Nature Switzerland, Cham, 2023.

\bibitem{mitra2022facilitating}
R.~Mitra, A.~Narechania, A.~Endert, and J.~Stasko.
\newblock Facilitating conversational interaction in natural language interfaces for visualization.
\newblock In {\em Proc. VIS}, pp. 6--10, 2022. \href{https://doi.org/10.1109/VIS54862.2022.00010}
{doi: {{%
10\hspace{.1pt}\discretionary{.}{%
}{.}\hspace{.4pt}1109\discretionary{/}{%
}{/}VIS54862\hspace{.1pt}\discretionary{.}{%
}{.}\hspace{.4pt}2022\hspace{.1pt}\discretionary{.}{%
}{.}\hspace{.4pt}00010}}}


\bibitem{naderifar2017snowball}
M.~Naderifar, H.~Goli, and F.~Ghaljaie.
\newblock Snowball sampling: A purposeful method of sampling in qualitative research.
\newblock {\em Strides in development of medical education}, 14(3), 2017.

\bibitem{nam2024using}
D.~Nam, A.~Macvean, V.~Hellendoorn, B.~Vasilescu, and B.~Myers.
\newblock Using an llm to help with code understanding.
\newblock In {\em 2024 IEEE/ACM 46th International Conference on Software Engineering (ICSE)}, pp. 881--881. IEEE Computer Society, 2024.

\bibitem{Narechania2020NL4DVAT}
A.~{Narechania}, A.~{Srinivasan}, and J.~{Stasko}.
\newblock {NL4DV}: A {Toolkit} for generating {Analytic Specifications} for {Data Visualization} from {Natural Language} queries.
\newblock {\em IEEE Trans. Vis. Comput. Graph.}, 2020. \href{https://doi.org/10.1109/TVCG.2020.3030378}
{doi: {{%
10\hspace{.1pt}\discretionary{.}{%
}{.}\hspace{.4pt}1109\discretionary{/}{%
}{/}TVCG\hspace{.1pt}\discretionary{.}{%
}{.}\hspace{.4pt}2020\hspace{.1pt}\discretionary{.}{%
}{.}\hspace{.4pt}3030378}}}


\bibitem{ngonga2013sorry}
A.-C. Ngonga~Ngomo, L.~B\"{u}hmann, C.~Unger, J.~Lehmann, and D.~Gerber.
\newblock Sorry, i don't speak sparql: translating sparql queries into natural language.
\newblock In {\em In Proc. ACM WWW}, WWW '13, p. 977–988. Association for Computing Machinery, New York, NY, USA, 2013. \href{https://doi.org/10.1145/2488388.2488473}
{doi: {{%
10\hspace{.1pt}\discretionary{.}{%
}{.}\hspace{.4pt}1145\discretionary{/}{%
}{/}2488388\hspace{.1pt}\discretionary{.}{%
}{.}\hspace{.4pt}2488473}}}


\bibitem{2024_unifying_llms_and_kgs}
S.~Pan, L.~Luo, Y.~Wang, C.~Chen, J.~Wang, and X.~Wu.
\newblock Unifying large language models and knowledge graphs: A roadmap.
\newblock {\em IEEE Transactions on Knowledge and Data Engineering}, pp. 1--20, 2024. \href{https://doi.org/10.1109/TKDE.2024.3352100}
{doi: {{%
10\hspace{.1pt}\discretionary{.}{%
}{.}\hspace{.4pt}1109\discretionary{/}{%
}{/}TKDE\hspace{.1pt}\discretionary{.}{%
}{.}\hspace{.4pt}2024\hspace{.1pt}\discretionary{.}{%
}{.}\hspace{.4pt}3352100}}}


\bibitem{petroni2019language}
F.~Petroni, T.~Rockt{\"a}schel, S.~Riedel, P.~Lewis, A.~Bakhtin, Y.~Wu, and A.~Miller.
\newblock Language models as knowledge bases?
\newblock In {\em Proc. EMNLP/IJCNLP}, pp. 2463--2473. ACL, Hong Kong, 2019. \href{https://doi.org/10.18653/v1/D19-1250}
{doi: {{%
10\hspace{.1pt}\discretionary{.}{%
}{.}\hspace{.4pt}18653\discretionary{/}{%
}{/}v1\discretionary{/}{%
}{/}D19\discretionary{%
}{-}{-}1250}}}


\bibitem{pirolli2005sensemaking}
P.~Pirolli and S.~Card.
\newblock The sensemaking process and leverage points for analyst technology as identified through cognitive task analysis.
\newblock In {\em In Proc. International Conf on Intelligence Analysis}, vol.~5, pp. 2--4. McLean, VA, USA, 2005.

\bibitem{rabiee2004focus}
F.~Rabiee.
\newblock Focus-group interview and data analysis.
\newblock {\em In Proc. NUTR SOC}, 63(4):655--660, 2004.

\bibitem{rapp2023collaborating}
A.~Rapp, A.~Boldi, L.~Curti, A.~Perrucci, and R.~Simeoni.
\newblock Collaborating with a text-based chatbot: An exploration of real-world collaboration strategies enacted during human-chatbot interactions.
\newblock In {\em In Proc. CHI 2023}, CHI '23. Association for Computing Machinery, New York, NY, USA, 2023. \href{https://doi.org/10.1145/3544548.3580995}
{doi: {{%
10\hspace{.1pt}\discretionary{.}{%
}{.}\hspace{.4pt}1145\discretionary{/}{%
}{/}3544548\hspace{.1pt}\discretionary{.}{%
}{.}\hspace{.4pt}3580995}}}


\bibitem{rawte2023survey}
V.~Rawte, A.~Sheth, and A.~Das.
\newblock A survey of hallucination in large foundation models.
\newblock {\em arXiv preprint arXiv:2309.05922}, 2023.

\bibitem{rozière2024code}
B.~Rozière, J.~Gehring, F.~Gloeckle, S.~Sootla, I.~Gat, X.~E. Tan, Y.~Adi, J.~Liu, R.~Sauvestre, T.~Remez, J.~Rapin, A.~Kozhevnikov, I.~Evtimov, J.~Bitton, M.~Bhatt, C.~C. Ferrer, A.~Grattafiori, W.~Xiong, A.~Défossez, J.~Copet, F.~Azhar, H.~Touvron, L.~Martin, N.~Usunier, T.~Scialom, and G.~Synnaeve.
\newblock Code llama: Open foundation models for code, 2024.

\bibitem{sambasivan2021everyone}
N.~Sambasivan, S.~Kapania, H.~Highfill, D.~Akrong, P.~Paritosh, and L.~M. Aroyo.
\newblock {“Everyone wants to do the model work, not the data work”}: {Data Cascades in High-Stakes AI}.
\newblock In {\em Proc. CHI}. ACM, New York, 2021. \href{https://doi.org/10.1145/3411764.3445518}
{doi: {{%
10\hspace{.1pt}\discretionary{.}{%
}{.}\hspace{.4pt}1145\discretionary{/}{%
}{/}3411764\hspace{.1pt}\discretionary{.}{%
}{.}\hspace{.4pt}3445518}}}


\bibitem{sen-etal-2023-knowledge}
P.~Sen, S.~Mavadia, and A.~Saffari.
\newblock Knowledge graph-augmented language models for complex question answering.
\newblock In B.~Dalvi~Mishra, G.~Durrett, P.~Jansen, D.~Neves~Ribeiro, and J.~Wei, eds., {\em In Proc. NLRSE}, pp. 1--8. Association for Computational Linguistics, Toronto, Canada, June 2023. \href{https://doi.org/10.18653/v1/2023.nlrse-1.1}
{doi: {{%
10\hspace{.1pt}\discretionary{.}{%
}{.}\hspace{.4pt}18653\discretionary{/}{%
}{/}v1\discretionary{/}{%
}{/}2023\hspace{.1pt}\discretionary{.}{%
}{.}\hspace{.4pt}nlrse\discretionary{%
}{-}{-}1\hspace{.1pt}\discretionary{.}{%
}{.}\hspace{.4pt}1}}}


\bibitem{setlur2016eviza}
V.~Setlur, S.~E. Battersby, M.~Tory, R.~Gossweiler, and A.~X. Chang.
\newblock Eviza: A natural language interface for visual analysis.
\newblock In {\em In Proc. ACM UIST}, pp. 365--377, 2016.

\bibitem{shen2022towards}
L.~Shen, E.~Shen, Y.~Luo, X.~Yang, X.~Hu, X.~Zhang, Z.~Tai, and J.~Wang.
\newblock Towards natural language interfaces for data visualization: A survey.
\newblock {\em IEEE transactions on visualization and computer graphics}, 2022.

\bibitem{shen2024dataplayer}
L.~Shen, Y.~Zhang, H.~Zhang, and Y.~Wang.
\newblock Data player: Automatic generation of data videos with narration-animation interplay.
\newblock {\em IEEE Transactions on Visualization and Computer Graphics}, 30(1):109--119, 2024. \href{https://doi.org/10.1109/TVCG.2023.3327197}
{doi: {{%
10\hspace{.1pt}\discretionary{.}{%
}{.}\hspace{.4pt}1109\discretionary{/}{%
}{/}TVCG\hspace{.1pt}\discretionary{.}{%
}{.}\hspace{.4pt}2023\hspace{.1pt}\discretionary{.}{%
}{.}\hspace{.4pt}3327197}}}


\bibitem{srinivasan2021snowy}
A.~Srinivasan and V.~Setlur.
\newblock Snowy: Recommending utterances for conversational visual analysis.
\newblock In {\em The 34th Annual ACM Symposium on User Interface Software and Technology}, pp. 864--880, 2021.

\bibitem{sultanum2023datatales}
N.~Sultanum and A.~Srinivasan.
\newblock Datatales: Investigating the use of large language models for authoring data-driven articles.
\newblock In {\em 2023 IEEE Visualization and Visual Analytics (VIS)}, pp. 231--235. IEEE, 2023.

\bibitem{sun2023think}
J.~Sun, C.~Xu, L.~Tang, S.~Wang, C.~Lin, Y.~Gong, H.-Y. Shum, and J.~Guo.
\newblock Think-on-graph: Deep and responsible reasoning of large language model with knowledge graph.
\newblock {\em arXiv preprint arXiv:2307.07697}, 2023.

\bibitem{tiddi2020knowledge}
I.~Tiddi, F.~L{\'{e}}cu{\'{e}}, and P.~Hitzler, eds.
\newblock {\em Knowledge Graphs for eXplainable Artificial Intelligence: Foundations, Applications and Challenges}, vol.~47 of {\em Studies on the Semantic Web}.
\newblock {IOS} Press, 2020.

\bibitem{toussaint2022troubles}
E.~Toussaint, P.~Guagliardo, L.~Libkin, and J.~Sequeda.
\newblock Troubles with nulls, views from the users.
\newblock {\em Proc. VLDB Endow.}, 15(11):2613–2625, 2022. \href{https://doi.org/10.14778/3551793.3551818}
{doi: {{%
10\hspace{.1pt}\discretionary{.}{%
}{.}\hspace{.4pt}14778\discretionary{/}{%
}{/}3551793\hspace{.1pt}\discretionary{.}{%
}{.}\hspace{.4pt}3551818}}}


\bibitem{tukey1977exploratory}
J.~W. Tukey et~al.
\newblock {\em Exploratory data analysis}, vol.~2.
\newblock Reading, MA, 1977.

\bibitem{vazquez2024llms}
P.-P. V{\'a}zquez.
\newblock Are llms ready for visualization?
\newblock {\em arXiv preprint arXiv:2403.06158}, 2024.

\bibitem{chainOfThought}
J.~Wei, X.~Wang, D.~Schuurmans, M.~Bosma, b.~ichter, F.~Xia, E.~Chi, Q.~V. Le, and D.~Zhou.
\newblock Chain-of-thought prompting elicits reasoning in large language models.
\newblock In S.~Koyejo, S.~Mohamed, A.~Agarwal, D.~Belgrave, K.~Cho, and A.~Oh, eds., {\em Advances in Neural Information Processing Systems}, vol.~35, pp. 24824--24837. Curran Associates, Inc., 2022.

\bibitem{wen2023mindmap}
Y.~Wen, Z.~Wang, and J.~Sun.
\newblock Mindmap: Knowledge graph prompting sparks graph of thoughts in large language models.
\newblock {\em arXiv preprint arXiv:2308.09729}, 2023.

\bibitem{wu2023autogen}
Q.~Wu, G.~Bansal, J.~Zhang, Y.~Wu, S.~Zhang, E.~Zhu, B.~Li, L.~Jiang, X.~Zhang, and C.~Wang.
\newblock Autogen: Enabling next-gen llm applications via multi-agent conversation framework.
\newblock {\em arXiv preprint arXiv:2308.08155}, 2023.

\bibitem{yang2023llm}
S.~Yang, M.~Teng, X.~Dong, and F.~Bo.
\newblock Llm-based sparql generation with selected schema from large scale knowledge base.
\newblock In {\em China Conference on Knowledge Graph and Semantic Computing}, pp. 304--316. Springer, 2023.

\bibitem{yao2019kg}
L.~Yao, C.~Mao, and Y.~Luo.
\newblock Kg-bert: Bert for knowledge graph completion.
\newblock {\em arXiv preprint arXiv:1909.03193}, 2019.

\bibitem{zeng2023large}
F.~Zeng, W.~Gan, Y.~Wang, N.~Liu, and P.~S. Yu.
\newblock Large language models for robotics: A survey.
\newblock {\em arXiv preprint arXiv:2311.07226}, 2023.

\bibitem{zhang2021neural}
J.~Zhang, B.~Chen, L.~Zhang, X.~Ke, and H.~Ding.
\newblock Neural, symbolic and neural-symbolic reasoning on knowledge graphs.
\newblock {\em AI Open}, 2:14--35, 2021.

\bibitem{zhang2023s}
Z.~Zhang, M.~Jia, B.~Yao, S.~Das, A.~Lerner, D.~Wang, T.~Li, et~al.
\newblock "it's a fair game'', or is it? examining how users navigate disclosure risks and benefits when using llm-based conversational agents.
\newblock {\em arXiv preprint arXiv:2309.11653}, 2023.

\bibitem{zheng2022telling}
C.~Zheng, D.~Wang, A.~Y. Wang, and X.~Ma.
\newblock Telling stories from computational notebooks: Ai-assisted presentation slides creation for presenting data science work.
\newblock In {\em In Proc. CHI 2022}, pp. 1--20, 2022.

\end{thebibliography}

\end{document}